\documentclass[10pt,aps,prb,twocolumn,noshowpacs,superscriptaddress]{revtex4}

\usepackage{mathrsfs, bm,color,amsmath,amssymb, stmaryrd, amsfonts,latexsym,graphicx}

\newcommand*{\dpar}[2]{\frac{\partial #1}{\partial #2}}
\newcommand*{\ddpar}[2]{\frac{\partial^2 #1}{\partial #2^2}}
\newcommand{\ud}{\,\mathrm{d}}
\newcommand{\Tr}{\mathrm{Tr}\,}
\newcommand{\tr}{\mathrm{tr}\,}
\newcommand{\nn}{\nonumber}
\newcommand{\be}{\begin{equation}}
\newcommand{\ee}{\end{equation}}

\begin{document}
\title{Orbital magnetism of coupled bands models}

\author{Arnaud \surname{Raoux}}
\email{arnaud.raoux@ens.fr}
\affiliation{Laboratoire de Physique des Solides, CNRS UMR 8502, Univ. Paris-Sud, F-91405 Orsay Cedex, France}
\affiliation{D\' epartement de Physique, \' Ecole Normale Sup\'erieure, 24 rue Lhomond, 75005 Paris, France}
\author{Fr\' ed\' eric \surname{Pi\' echon}}
\affiliation{Laboratoire de Physique des Solides, CNRS UMR 8502, Univ. Paris-Sud, F-91405 Orsay Cedex, France}
\author{Jean-No\"el \surname{Fuchs}}
\affiliation{Laboratoire de Physique des Solides, CNRS UMR 8502, Univ. Paris-Sud, F-91405 Orsay Cedex, France}
\affiliation{Laboratoire de Physique Th\' eorique de la Mati\` ere Condens\' ee, CNRS UMR 7600, Univ. Pierre et Marie Curie 4, place Jussieu, 75252 Paris Cedex 05, France}
\author{Gilles \surname{Montambaux}}
\affiliation{Laboratoire de Physique des Solides, CNRS UMR 8502, Univ. Paris-Sud, F-91405 Orsay Cedex, France}

\date{\today}

\begin{abstract}
We develop a gauge-independent perturbation theory for the grand potential of itinerant electrons in two-dimensional tight-binding models in the presence of a perpendicular magnetic field. At first order in the field, we recover the result of the so-called {\it modern theory of orbital magnetization} and, at second order, deduce a new general formula for the orbital susceptibility. In the special case of two coupled bands, we relate the susceptibility to geometrical quantities such as the Berry curvature. Our results are applied to several two-band -- either gapless or gapped -- systems. We point out some surprising features in the orbital susceptibility -- such as in-gap diamagnetism or parabolic band edge paramagnetism -- coming from interband coupling. From that we draw general conclusions on the orbital magnetism of itinerant electrons in multi-band tight-binding models.
\end{abstract}


\maketitle
\section{Introduction}

The magnetic response of itinerant electronic systems in the absence of spin-orbit coupling can be split in two different parts: spin and orbital contributions. The spin susceptibility is easily understood in terms of Pauli paramagnetism as it only depends on the density of states at the Fermi level\cite{Ashcroft}. There is no essential difference between the case of free electrons and that of Bloch electrons. In the following, we therefore consider spinless electrons and focus on orbital magnetism of itinerant electrons. The study of the later begun with Landau~\cite{Landau30} for free electrons, and continued with Peierls~\cite{Peierls33} who took explicitly the effect of the periodic potential into account: he derived a formula which is valid in a one-band approximation (single band tight-binding model). After these pioneering works, a lot of effort has been put in trying to generalize the so-called Landau-Peierls (LP) formula to many-band systems~\cite{Adams53,Hebborn60,Roth62,Wannier64,Misra69} with different approaches: effective multiband Hamiltonians, use of Bloch or Wannier functions,~etc. The challenge was to tackle the case of coupled bands, the contribution of which cannot be treated separately. However, the resulting formulae were so complicated that any attempt of physical interpretation was vain, and a complete evaluation was in general impossible. Fukuyama~\cite{Fukuyama71} first gave a very compact expression for the susceptibility in terms of Green's functions using a slowly varying vector potential for a perfect periodic system. While his linear response formula gave interesting results (for example in bismuth\cite{Fukuyama70}), it seems to be incomplete for tight-binding systems as it does not recover the LP formula in the single-band limit. In the context of graphene, the Fukuyama formula was recently completed for the tight-binding model by Gomez-Santos and Stauber~\cite{Gomez-Santos11} and anticipated by Koshino and Ando~\cite{Koshino07}.

Graphene, first theoretically studied by Wallace~\cite{Wallace47} and experimentally discovered sixty years later by Novoselov and Geim~\cite{Novoselov04}, is a honeycomb lattice of carbon atoms that remains conducting despite its minimal thickness. It is essentially a strongly coupled two-band system that is ideally suited to test the prediction of orbital susceptibility formulae. The simplest tight-binding model describing graphene~\cite{Wallace47} can actually be considered as a paradigmatic case of strong band coupling. Like bismuth, graphite is known for its huge diamagnetism which seems to be also experimentally observed in graphene~\cite{Sepioni10,Nikolaev13}. McClure~\cite{McClure56} derived such a property from graphene's unusual Landau levels at half filling. He showed that, when the chemical potential $\mu$ is right at the Dirac point (usually at $\mu=0$), the susceptibility becomes infinitely diamagnetic as the temperature vanishes. This can not be recovered by the LP formula and is therefore a signature of interband effects on the magnetic response of graphene. However, McClure's formula predicts a null susceptibility as soon as $\mu\not=0$. This is not correct for at least two reasons. First, it violates an exact sumrule (see below). Second, Vignale~\cite{Vignale91} showed quite generally that, in the vicinity of a saddle point in the dispersion relation (corresponding to a van Hove singularity in the density of states, which occurs in graphene at finite energy $\sim \pm 3$~eV), the susceptibility should actually be infinitely paramagnetic and not zero. The Fukuyama formula\cite{Fukuyama71} correctly recovers the McClure diamagnetic peak\cite{Fukuyama07} and the van Hove paramagnetism but fails to describe the exact chemical potential dependance of the tight-binding model, as we discuss below. The formula derived by Ref.~[\onlinecite{Gomez-Santos11}] succeeds in giving the susceptibility of graphene for any value of the chemical potential (controlling the band filling) and a numerical approach done in Ref.~[\onlinecite{Raoux14}] based on the energy spectrum of the honeycomb lattice in a magnetic field (graphene's Hofstadter butterfly~\cite{Rammal85}) confirmed it. 

In order to derive the orbital susceptibility, the authors of Ref.~[\onlinecite{Gomez-Santos11}] use a gauge-dependent procedure in which they employ a trick to derive a continuous version of the tight-binding Hamiltonian. In particular, the derivation of the effective current operator seems ambiguous since different non-equivalent expressions can be found (even if the zero wavevector limit remains the same). Several recent works~\cite{Savoie12, Chen11,Nourafkan14,Swiecicki14} derived a gauge-independent perturbation theory of the grand potential. Refs.~[\onlinecite{Chen11,Nourafkan14}] restrict to the magnetization, and Ref.~[\onlinecite{Savoie12}] gives a (rather elaborate) formula for the orbital susceptibility.

This paper presents a perturbation theory for independent particles (Sec.~\ref{sec:derivation}) in terms of the magnetic field using Green's functions which are explicitly gauge-independent~\cite{Savoie12}; the derivation presents a straightforward physical interpretation as it does not use a continuous limit. In addition, this method easily allows one to get both the magnetization -- related to the first-order term in the magnetic field -- and the orbital susceptibility as the second-order term. The expansion actually holds to any order in the magnetic field. In particular we recover (Sec.~\ref{sec:magnetization}) the formula of the magnetization in terms of the Berry curvature and the orbital magnetic moment, as obtained in the so-called \emph{modern theory of orbital magnetization}~\cite{Thonhauser11,Thonhauser05,Xiao05}. 
In Sec.~\ref{sec:susceptibility}, we present a new formula for the orbital susceptibility, see Eq.~(\ref{eq:chi}), and compare it to the different results listed in the introduction. In particular, it agrees with the formula of Ref.~[\onlinecite{Gomez-Santos11}]. Next, we derive a convenient formula for the orbital susceptibility of two-band tight-binding models (see Eq.~(\ref{eq:chi_deuxbandes})), that we apply to several specific Hamiltonians (Sec.~\ref{sec:two-band}) in order to gain insight on the importance of interband coupling. Equations~(\ref{eq:chi}) and (\ref{eq:chi_deuxbandes}) are the main results of this paper. Eventually, we give a general conclusion on orbital magnetism of coupled bands models in Sec.~\ref{sec:conclusion}.

\section{General derivation}
\label{sec:derivation}
\paragraph*{Motivation---} We restrict ourselves to 2D systems to simplify the algebra.
In the presence of a magnetic field $B$ along the transverse axis $z$, the grand canonical potential of a Fermi-Dirac gas of non-interacting electrons yields
\begin{equation}
\label{eq:def_GP}
\Omega(T,\mu,B)=-T\int_{-\infty}^{+\infty} \ln\left(1+e^{-(E-\mu)/T}\right)\rho(E,B)\,\mathrm{d}E
\end{equation}
where $\rho(E,B)$ is the density of states (DoS) of the system (we use units such that the Boltzmann constant $k_\mathrm{B}=1$). The quantities of interest are derivatives of the grand potential taken in the limit $B=0$: the magnetization
\begin{equation}
\label{eq:def_M}
M(\mu,T)=-\frac{1}{S}\left.\dpar{\Omega}{B}\right|_{B=0}
\end{equation}
and the orbital susceptibility ($\mu_0=4\pi\cdot10^{-7}$ in S.I. units)
\begin{equation}
\label{eq:def_chi}
\chi_\mathrm{orb}(\mu,T)=-\frac{\mu_0}{S}\left.\ddpar{\Omega}{B}\right|_{B=0}
\end{equation}
which, via $\Omega$, only depend on derivatives of $\rho(E,B)$. $S$ is the sample area. In 2D, $\chi_\mathrm{orb}$ is homogeneous to a length. The DoS can be written in terms of the retarded Green's function $G$ of the system
\begin{equation}
\label{eq:def_g}
\rho(E,B)=-\frac{1}{\pi}\Im m \Tr\,G(E,B)
\end{equation}
so that we will search for a perturbation theory of $G$.

A useful tool in the following will be the magnetic sumrule: it can be shown~(see Supplemental material of Ref.~[\onlinecite{Raoux14}]) that
\begin{equation}
\label{eq:sumrule}
\frac{\partial}{\partial B}\int_{-\infty}^{+\infty}\Omega(T,\mu,B)\ud\mu=0
\end{equation}
from which we deduce the sumrule relative to the susceptibility
\begin{equation}
\int_{-\infty}^{+\infty}\chi_\mathrm{orb}(\mu,T)\ud\mu=0.
\end{equation}
This sumrule holds for any tight-binding model, provided the magnetic field only enters as a Peierls phase on the hopping amplitudes in the Hamiltonian (see below).

\begin{figure}[t]
\begin{center}
\includegraphics[scale=0.9]{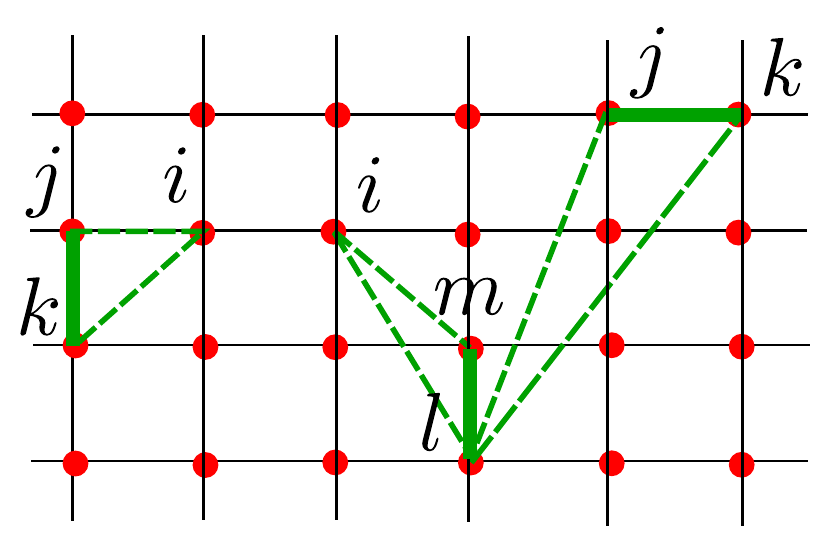}
\caption{(Color online). $\Phi_{jki}$ is a dimensionless magnetic flux corresponding to the area of a triangle made of one side $jk$ associated to a hopping amplitude $t_{jk}$ and one site $i$ closing the triangle. It serves to illustrate Eq.~(\ref{eq:Gii}). Likewise $\Phi_{jkl}$ and $\Phi_{lmi}$ illustrating Eq.~(\ref{eq:2nd_order_1b}) are also plotted.}
\label{fig:phases}
\end{center}
\end{figure}

\paragraph*{System---} Starting from a tight-binding Hamiltonian in real-space representation in the absence of a magnetic field:
\begin{equation}
h=\sum_{i,j}t_{ij}\left|i\right\rangle\left\langle j\right|
\end{equation} 
where $i$ and $j$ are site indices, the magnetic field is taken into account by performing the Peierls substitution $t_{ij}\to t_{ij}e^{i\varphi_{ij}}$ where
\begin{equation}
\varphi_{ij}=\frac{e}{\hbar}\int_i^j\bm A(\bm r)\cdot \mathrm{d}\bm{l},
\end{equation}
$\bm A(\bm r)$ being a vector potential corresponding to a static uniform magnetic field. Let us define the quantity
\begin{equation}
\Phi_{ijk}=\varphi_{ij}+\varphi_{jk}+\varphi_{ki}=\frac{e}{\hbar}\oint_{ijk}\bm A(\bm r)\cdot \mathrm{d}\bm{l}=\frac{e}{\hbar}\iint_{ijk}\bm B\cdot\mathrm{d}\bm{S}.
\end{equation}
If $\varphi_{ij}$ depends on the gauge, $\Phi_{ijk}$ is gauge-independent using Stokes' formula since it is proportional to the magnetic flux through the oriented triangle $(ijk)$ (cf.~Fig.~\ref{fig:phases}). Its explicit expression in Cartesian coordinates is
%

%
\begin{equation}
\label{eq:Phi}
\Phi_{ijk}=\frac{eB}{2\hbar}\left[(x_i-x_k)(y_k-y_j)-(y_i-y_k)(x_k-x_j)\right].
\end{equation}

\paragraph*{Perturbation theory---}

The total Hamiltonian in a $B$ field reads
\begin{equation}
H=\sum_{ij}t_{ij}e^{i\varphi_{ij}}\left|i\right\rangle\left\langle j\right|.
\end{equation} 
Let $G(E,B)$ (resp. $g(E)$) be the Green's function relative to $H$ (resp. $h$). If one directly expands $G$ in powers of $g$, one gets a gauge-dependent expression. A trick\cite{Savoie12, Chen11} to circumvent such a problem is to define a new ``twisted'' Green's function $\tilde g$ by
\begin{equation}
\tilde g_{ij}=e^{i\varphi_{ij}}g_{ij},
\end{equation}
then expand $G$ in terms of $\tilde g$ and finally reintroduce $g$ in order to recover a gauge-independent expression for the diagonal elements of $G$.

One easily shows that
\begin{equation}
\label{eq:G}
(E-H)\tilde g=1-\mathcal{T}
\end{equation}
with
\begin{equation}
\mathcal{T}_{ij}=e^{i\varphi_{ij}}\sum_k\left(e^{i\Phi_{ikj}}-1\right)t_{ik}g_{kj}.
\end{equation}
Eq.(\ref{eq:G}) gives:
\begin{equation}
\label{eq:expression_G}
G=(E-H)^{-1}=\tilde g(1-\mathcal{T})^{-1}=\tilde g\sum_{n\geq0}\mathcal{T}^n.
\end{equation}
Because the interesting part is the trace of $G$, only the diagonal terms in Eq.~(\ref{eq:expression_G}) need to be considered. Let $G^{(1)}$ and $G^{(2)}$ denote the linear and quadratic terms in the magnetic field respectively. The expansion of Eq.~(\ref{eq:expression_G}) in $B$ gives
\begin{equation}
\label{eq:Gii}
G_{ii}^{(1)}=i\sum_{jk}\Phi_{jki}\, g_{ij}\, t_{jk}\, g_{ki}
\end{equation}
and
\begin{subequations}\label{eq:2nd_order_1}
\begin{align}
G_{ii}^{(2)}=&-\frac{1}{2}\sum_{jk}\Phi_{jki}^2\, g_{ij}\, t_{jk}\, g_{ki}\label{eq:2nd_order_1a}\\
&-\sum_{jklm}\Phi_{jkl}\Phi_{lmi}\, g_{ij}\, t_{jk}\, g_{kl}\, t_{lm}\, g_{mi}\label{eq:2nd_order_1b}.
\end{align}
\end{subequations}
The diagonal quantities $G^{(m)}_{ii}$ ($m\in\mathbb N$) only depend on gauge-independent variables; it is not the case of off-diagonal elements. Note that second-order expansion in $B$ contains first and second powers in the operator $\mathcal{T}$. The two terms Eqs.~(\ref{eq:2nd_order_1a}) and~(\ref{eq:2nd_order_1b}) are of order $B^2$ and are reminiscent of the Larmor (first order in $\mathcal{T}$) and Van Vleck (second order in $\mathcal{T}$) contributions, that are well-known in the magnetism of isolated atoms\cite{Ashcroft}

Using Eq.~(\ref{eq:Phi}) and defining position operators \mbox{$x=\sum_i x_i\left|i\right\rangle\left\langle i\right|$} and $y=\sum_i y_i\left|i\right\rangle\left\langle i\right|$, $G^{(1)}_{ii}$ and $G^{(2)}_{ii}$ can be expressed in terms of commutators of $h$, $g$, $x$ and $y$. To shorten the notations, if $O$ is an operator, $-\frac{i}{\hbar}[x,O]$ (resp.~$-\frac{i}{\hbar}[y,O]$) will be denoted $O^x$ (resp. $O^y$); this notation will correspond to a derivative with respect to $k_x$ (resp. $k_y$) in the $\bm k$-space representation. The particular case where $O$ is the Hamiltonian yields the velocity operator $v_x=\dot x=-\frac{i}{\hbar}[x,H]$. With these notations, Eqs.~(\ref{eq:Gii}) and~(\ref{eq:perturbation_2nd_order}) become
\begin{equation}
\label{eq:perturbation_1st_order}
G^{(1)}_{ii}=-\frac{ieB}{2\hbar}\left(gh^xg^y- gh^yg^x\right)_{ii}
\end{equation}
and 
\begin{multline}
\label{eq:perturbation_2nd_order}
G^{(2)}_{ii}= -\frac{e^2B^2}{8\hbar^2}\Big[g\Big(h^{xx}g^{yy}+h^{yy}g^{xx}-2h^{xy}g^{xy}\\ \left.\left.
+2\left(h^xg^y-h^yg^x\right)^2\right)\right]_{ii}.
\end{multline}
The last term of Eq.~(\ref{eq:perturbation_2nd_order}) looks like the square of the first order correction. It corresponds to the Van Vleck term of atomic physics. The equality $g^x=-\frac{i}{\hbar}[x,g]=gh^xg$ (and equivalently with $y$) will be useful to compute the commutators of the Green's functions.

\section{Orbital Magnetization}
\label{sec:magnetization}

In a perfect crystal, the set of Bloch functions diagonalizes the Hamiltonian. If $\psi_{n\bm k}(\bm r)=e^{i\bm k\cdot\bm r}u_{ n\bm k}(\bm r)$ is a Bloch function, eigenfunction of the Hamiltonian $h$ with eigenvalue $\varepsilon_{n\bm k}$, then $u_{n\bm k}$ is a periodic function with the periodicity of the lattice, and is an eigenfunction of $h_{\bm k}=e^{-i\bm r\cdot\bm k}he^{i\bm r\cdot\bm k}$ with the same eigenvalue $\varepsilon_{n\bm k}$ ($\bm r$ being the position operator in the definition of $h_{\bm k}$).
Using this basis, it is a little long but straightforward (cf. Appendix~\ref{appendix:magnetization}) to recover from Eqs.~(\ref{eq:def_GP}, \ref{eq:def_M}, \ref{eq:def_g}, \ref{eq:perturbation_1st_order}) the nowadays well-established formula\cite{Xiao05,Thonhauser05} for the magnetization 
\begin{multline}
\label{eq:M}
M=\sum_n\int_\mathrm{BZ}\left[n_\mathrm{F}(\varepsilon_{n\bm k})m_{n\bm k}\right.\\
\left. +\frac{eT}{\hbar}\ln\left(1+e^{-\beta(\varepsilon_{n\bm k}-\mu)}\right)\Omega_{n\bm k}\right]\frac{\ud^2k}{4\pi^2}
\end{multline}
in terms of the Berry curvature $\Omega_{n\bm k}$ and the magnetic moment $m_{n\bm k}$ along the $z$ axis\cite{Chen11,Nourafkan14}:
\begin{align}
\Omega_{n\bm k}&=i\left\langle \partial_{\bm k}u_{n\bm k}\right|\times\left|\partial_{\bm k}u_{n\bm k}\right\rangle\cdot\bm u_z\nn \\
&=i\sum_{m\not=n}\left\langle u_{n\bm k}\left|\frac{h_{\bm k}^x\mathcal P_{m\bm k}h_{\bm k}^y-h_{\bm k}^y\mathcal P_{m\bm k}h_{\bm k}^x}{(\varepsilon_{n\bm k}-\varepsilon_{m\bm k})^2}\right|u_{n\bm k}\right\rangle\label{eq:def_berry} \\
m_{n\bm k}&=-\frac{ie}{2\hbar}\left\langle \partial_{\bm k}u_{n\bm k}\right|\times(h_{\bm k}-\varepsilon_{n\bm k})\left|\partial_{\bm k}u_{n\bm k}\right\rangle\cdot\bm u_z\nn \\
&=\frac{ie}{2\hbar}\sum_{m\not=n}\left\langle u_{n\bm k}\left|\frac{h_{\bm k}^x\mathcal P_{m\bm k}h_{\bm k}^y-h_{\bm k}^y\mathcal P_{m\bm k}h_{\bm k}^x}{\varepsilon_{n\bm k}-\varepsilon_{m\bm k}}\right|u_{n\bm k}\right\rangle,
\label{eq:def_orbital}
\end{align}
where $n_\mathrm{F}(E)=\left(1+e^{(E-\mu)/T}\right)^{-1}$  is the Fermi function (implicitely depending on $\mu$ and $T$) and $\mathcal P_{n\bm k}$ the projector from the Hilbert space to the state $|u_{n\bm k}\rangle$.

The above derivation can be straightforwardly extended to the case of a finite and disordered system described by a Hamiltonian $h$. If we call  $\{|\psi_\alpha\rangle,\varepsilon_\alpha\}$ a set of eigenstates and eigenvalues of $h$, then equations corresponding to (\ref{eq:M}), (\ref{eq:def_berry}) and (\ref{eq:def_orbital}) are obtained throught the substitutions $h_{\bm k}\to h$, $|u_{n\bm k}\rangle \to |\psi_\alpha\rangle$ and $\varepsilon_{n, \bm k}\to \varepsilon_\alpha$. In addition, starting from Eq.~(\ref{eq:perturbation_1st_order}), one could also derive the local orbital magnetization (see, e.g. [\onlinecite{Bianco13}]).

%
%
%
%

In a system that is time-reversal invariant, the spontaneous magnetization vanishes and one needs to go to the second order response in order to obtain orbital magnetism.

\section{Orbital Susceptibility}
\label{sec:susceptibility}

We now derive a new formula for the orbital susceptibility. Starting from Eq.~(\ref{eq:perturbation_2nd_order}) and after some algebra (cf. Appendix \ref{appendix:2nd_order}), one obtains:
%
\begin{multline}
\Tr G^{(2)}=\frac{e^2B^2}{24\hbar^2}\dpar{}{E}\Tr\big[h^{xx}gh^{yy}g-h^{xy}gh^{xy}g\\
-4\left(gh^xgh^xgh^ygh^y-gh^xgh^ygh^xgh^y\right)\big].
\end{multline}
$h^{i}$ should be understood as the $k_i$-derivative of $h_{\bm k}$, $i=x,y$. The general formula for the orbital susceptibility follows:
\begin{widetext}
\begin{equation}
\label{eq:chi}
\chi_\mathrm{orb}(\mu,T)=-\frac{\mu_0 e^2}{12\hbar^2}\frac{\Im m}{\pi S}\int_{-\infty}^{+\infty} n_\mathrm{F}(E)\Tr\left\{gh^{xx}gh^{yy}-gh^{xy}gh^{xy}- 4(gh^xgh^xgh^ygh^y-gh^xgh^ygh^xgh^y)\right\}\ud E.
\end{equation}
%
Equation~(\ref{eq:chi}) is the first main result of this paper. As for the orbital magnetization, the above formula is valid even if the system does not have translational symmetry such as molecules, ribbons or disordered systems. In the remaining of the present paper, we restrict to infinite crystals.

In the case of a single band, $\Tr(\bullet)=\sum_{\bm k}=S\int_\mathrm{BZ} \frac{\ud^2k}{4\pi^2}$ where the integration is performed over the first Brillouin zone (BZ), the last term (in parenthesis) in the trace of this formula vanishes and one immediately recovers the Peierls formula\cite{Peierls33}
\begin{equation}
\label{eq:chi_LP}
\chi_\mathrm{orb}(\mu,T)=\frac{\mu_0 e^2}{12\hbar^2}\int_\mathrm{BZ} n'_\mathrm{F}(\varepsilon_{\bm k})\left(\varepsilon^{xx}_{\bm k}\varepsilon^{yy}_{\bm k}-(\varepsilon^{xy}_{\bm k})^2\right)\frac{\ud^2k}{4\pi^2}
\end{equation}
with $\varepsilon_{\bm k}$ the energy spectrum. A detailed discussion on the use of the one-band LP formula is provided in Appendix~\ref{appendix:LP_square}. Moreover, using a partial integration in Eq.~(\ref{eq:chi}), we recover another expression of the susceptibility obtained in Ref.~[\onlinecite{Gomez-Santos11}](Eq.~(3)):
\begin{equation}
\label{eq:chi_Stauber}
\chi_\mathrm{orb}(\mu,T)=-\frac{\mu_0 e^2}{2\hbar^2}\frac{\Im m}{\pi S}\int_{-\infty}^{+\infty} n_\mathrm{F}(E)\Tr\left\{gh^{x}gh^{y}gh^{x}gh^{y}+ \frac{1}{2}(gh^xgh^y+gh^ygh^x)gh^{xy}\right\}\ud E.
\end{equation}
\end{widetext}

%
%
Fukuyama's formula\cite{Fukuyama71} is the first term in the trace of Eq.~(\ref{eq:chi_Stauber}). 
Nevertheless, Fukuyama's formula does not recover the LP result in the one-band case (see Fukuyama's discussion of that point in Ref.~[\onlinecite{Fukuyama71}] and our discussion in Appendix~\ref{appendix:LP_square}). Actually, the Fukuyama formula does not work for tight-binding models that are not separable, \emph{i.e.} such that $h^{xy}\neq 0$, see [\onlinecite{Gomez-Santos11,Koshino07}]. In order to recover the LP formula, one needs to consider the full Eq.~(\ref{eq:chi_Stauber}), as the two terms contribute in the one-band limit. In this regard, Eq.~(\ref{eq:chi}) is more adequate than Eq.~(\ref{eq:chi_Stauber}) to such a comparison: indeed, the first two terms (quadratic in $g$) of the trace in Eq.~(\ref{eq:chi}) consist in a ``generalized'' version of Eq.~(\ref{eq:chi_LP}), and if $h^x$ and $h^y$ commute with $h$ (a one-band Hamiltonian is a scalar), the last term vanish, and this directly leads to the LP formula. For this reason, Eq.~(\ref{eq:chi}) is preferred in the following.

The LP result is strictly valid only for a single band. However, it does contain interesting physics that is useful even when discussing two-band models with band coupling (see Appendix~\ref{appendix:LP_square}). In the multi-band case, the LP formula can be trivially extended to an approximate ``band by band'' formula\cite{Fukuyama71}
\begin{equation}
\chi_\mathrm{LP}=\frac{\mu_0e^2}{12\hbar^2}\sum_n \int_\mathrm{BZ} n'_\mathrm{F}(\varepsilon_{n\bm k})\left(\varepsilon_{n, \bm k}^{xx}\varepsilon_{n, \bm k}^{yy}-(\varepsilon_{n, \bm k}^{xy})^2\right)\frac{\ud^2k}{4\pi^2}\, ,
\end{equation}
which neglects all interband effects. It will serve for comparison purposes below.


\section{Application to two-band models}
\label{sec:two-band}


\subsection{Two-band formula}

In this section, we derive a formula valid for two-band models with particle-hole symmetry, in order to illustrate as clearly as possible the effects of interband coupling on the magnetic response of a crystal. 

For such models, the $\bm k$-space Hamiltonian matrix can be written $h_{\bm k}=\bm f_{\bm k}\cdot \bm\sigma$ where $\bm\sigma$ is the vector of Pauli matrices, and $\bm f_{\bm k}$ a 3-dimensional vector depending on the 2-dimensional vector $\bm k$. This Hamiltonian has two eigenvalues for each $\bm k$: $\varepsilon_{s\bm k}=s\varepsilon_{\bm k}=s|\bm f_{\bm k}|$ with $s=\pm$. They are associated to two eigenspaces defined by their projectors: $\mathcal P_{s\bm k}=\frac12(1+s\bm\sigma\cdot\bm n_{\bm k})$, where $\bm n_{\bm k}=\bm f_{\bm k}/\varepsilon_{\bm k}$ is a unit vector on the sphere $\mathcal S^2$. The Hamiltonian reads $h_{\bm k}=\varepsilon_{\bm k}\bm n_{\bm k}\cdot \bm\sigma$. In this basis, $h_{\bm k}$ and $g_{\bm k}(E)=(E-h_{\bm k})^{-1}$ are diagonal and in particular: $g_{\bm k}(E)=\sum_sg_{s\bm k}(E)\mathcal P_{s\bm k}$ with $g_{s\bm k}(E)=(E-\varepsilon_{s\bm k})^{-1}$. Interband coupling arises because derivatives of the Hamiltonian are not diagonal in this basis. For brevity, the $\bm k$-dependence will be implicit in the following except in definitions.

In Eq.~(\ref{eq:chi}), the trace is $\Tr(\bullet)=\sum_{\bm k}\mathrm{tr}(\bullet)=S\int \frac{\ud^2k}{4\pi^2}\mathrm{tr}(\bullet)$ where the integration is performed over the first Brillouin zone (BZ), $\mathrm{tr}(\bullet)$ being the partial trace operator on the band index. We separate two different contributions: 
\begin{align}
U_{\bm k}(E)&=\tr\left\{ (g h^{xx} g h^{yy}- g h^{xy} g h^{xy}\right )_{\bm k}\} \\
V_{\bm k}(E)&=\tr\left\{(g h^x g h^x g h^y g h^y-g h^x g h^y g h^x g h^y)_{\bm k}\right\}\, .
\end{align}
$U$ and $V$ qualitatively differ because $U$ is made of second-order derivatives of the Hamiltonian, but $V$ is only composed of first-order derivatives. 

After some algebra (see Appendix \ref{appendix:2band} for details), 
the susceptibility can be written:
%
\begin{widetext}
\begin{equation}
\label{eq:chi_deuxbandes}
\chi_\mathrm{orb}(\mu,T)=\frac{\mu_0 e^2}{12\hbar^2}\sum_{s=\pm} \int_\mathrm{BZ}\left[(U_1-V_1-4V_2)\left(n_F' -\frac{s n_F}{\varepsilon_{\bm k}}\right)+U_2 \frac{s n_F}{\varepsilon_{\bm k}}-V_1\varepsilon_{\bm k} n_F''\right]\frac{\ud^2k}{4\pi^2}
\end{equation}
\end{widetext}
%
with $n_F$ a shorthand notation for $n_F(\varepsilon_{s \bm k})$ and $n_F'$, $n_F''$ are first and second derivatives of $n_F$. We have also defined the quantities:
\begin{align}
\varepsilon^2 U_{1\bm k}&=(\bm f^{xx} \cdot \bm f)(\bm f^{yy} \cdot \bm f)-(\bm f^{xy} \cdot \bm f)^2\nn \\
U_{2\bm k}&=\bm f^{xx}\cdot \bm f^{yy}-\bm f^{xy}\cdot\bm f^{xy}\nn \\
\varepsilon^2 V_{1\bm k}&=\left(\varepsilon^y\bm f^x-\varepsilon^x\bm f^y\right)^2\nn \\
\varepsilon^4 V_{2\bm k}&=(\left(\bm f^x\times\bm f^y\right)\cdot\bm f)^2=4\varepsilon^6 \Omega_{s\bm k}^2\, 
\end{align}
with $\varepsilon\equiv \varepsilon_{\bm k}$ and $\bm f \equiv \bm f_{\bm k}$. The last quantity has been written in terms of the Berry curvature $\Omega_{s \bm k}=\frac{s}{2}({\bm n}^x\times {\bm n}^y)\cdot {\bm n}$, which, in the particle-hole symmetric two-band case, is also related to the orbital magnetic moment $m_{\bm k}=\frac{e}{\hbar}\varepsilon_{s\bm k} \Omega_{s\bm k}$ \cite{Xiao10,Fuchs10}. We have used this expression to compute the susceptibility in various examples presented below.

%


Eq.~(\ref{eq:chi_deuxbandes}) is the second main result of the present paper. It is a computable expression for numerical integration and is the starting point to discuss some examples in the next sections. 
Furthermore, the way it is written, each three parts verifies independently the susceptibility sumrule in Eq.~(\ref{eq:sumrule}).
However, it is also possible to have expressions in which only $n_F$ and $n_F'$ appear (and no longer $n_F''$). This work will be presented elsewhere\cite{RaouxUnpublished}. It is tempting to interpret terms proportional to $n_F'$ as Fermi surface contributions and those proportional to $n_F$ as bulk Fermi sea contributions.


In the remaining part of the paper, we apply Eq.~(\ref{eq:chi_deuxbandes}) to different two-band models in order to highlight some features in the orbital magnetic response linked to interband coupling.

\subsection{Gapless Systems}

\begin{figure}[t]
\begin{center}
\includegraphics[scale=0.6]{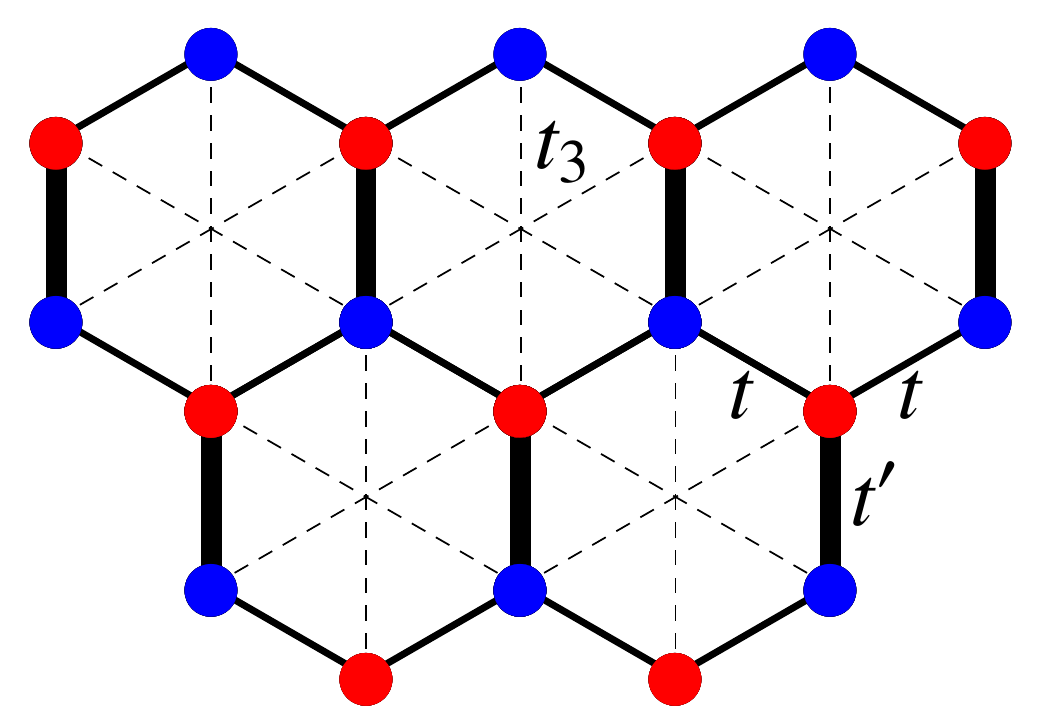}
\caption{(color online) Honeycomb lattice with first $t$ (full lines) and third $t_3$ (dashed lines) nearest-neighbor hopping amplitudes. The two sites in the unit cell are called $A$ and $B$ and are shown as blue and red dots. In a uniaxially compressed honeycomb lattice, there are two different values for the nearest neighbor hopping amplitudes: $t$ for thin (non--vertical) lines and $t'\geq t$ for thick (vertical) lines.}
\label{fig:schema}
\end{center}
\end{figure}

\subsubsection{Graphene} 
The honeycomb lattice is plotted in Fig.~\ref{fig:schema}. We consider the usual model of graphene, \emph{i.e.} the nearest-neighbor tight-binding model with hopping amplitude $t$ (corresponding to $t'=t$ and $t_3=0$ in Fig.~\ref{fig:schema}). The $\bm k$-space Hamiltonian can be written as $h_{\bm k}=\bm f_\mathrm{gr}\cdot \bm \sigma$ with\cite{CastroNeto09}
\begin{align}
\nonumber
\bm f_\mathrm{gr}&=t\sum_{1\leq j\leq3}\left(
\begin{array}{c}
\cos(\bm k\cdot\bm \delta_i) \\
\sin(\bm k\cdot\bm \delta_i) \\
0
\end{array}
\right)\\ \label{eq:f_graphene}
&=t\left(
\begin{array}{c}
\cos(k_y)+2\cos(k_y/2)\cos(\sqrt3k_x/2) \\
-\sin(k_y)+2\sin(k_y/2)\cos(\sqrt3k_x/2) \\
0
\end{array}
\right),
\end{align}
in which $\bm\delta_i$ ($1\leq i\leq3$) are the vectors linking a $A$~atom to its three $B$ nearest-neighbors (Fig.~\ref{fig:schema}). From now on, we use units such that the nearest neighbor distance $a=1$, $t=1$ and $\hbar=1$. We also introduce a convenient susceptibility scale $\chi_0\equiv \mu_0 e^2 t a^2/\hbar^2=\mu_0 e^2$ (with typical values $t\sim 1$~eV and $a\sim 1$~\AA, $\chi_0\sim 10^{-5}$~\AA \, corresponding in 3D to $\chi_0^{3D}\sim \chi_0/a \sim 10^{-5}$. This is as large as orbital susceptibility gets in experiments, apart from superconductors).
\begin{figure}[t]
\begin{center}
\includegraphics[scale=0.41]{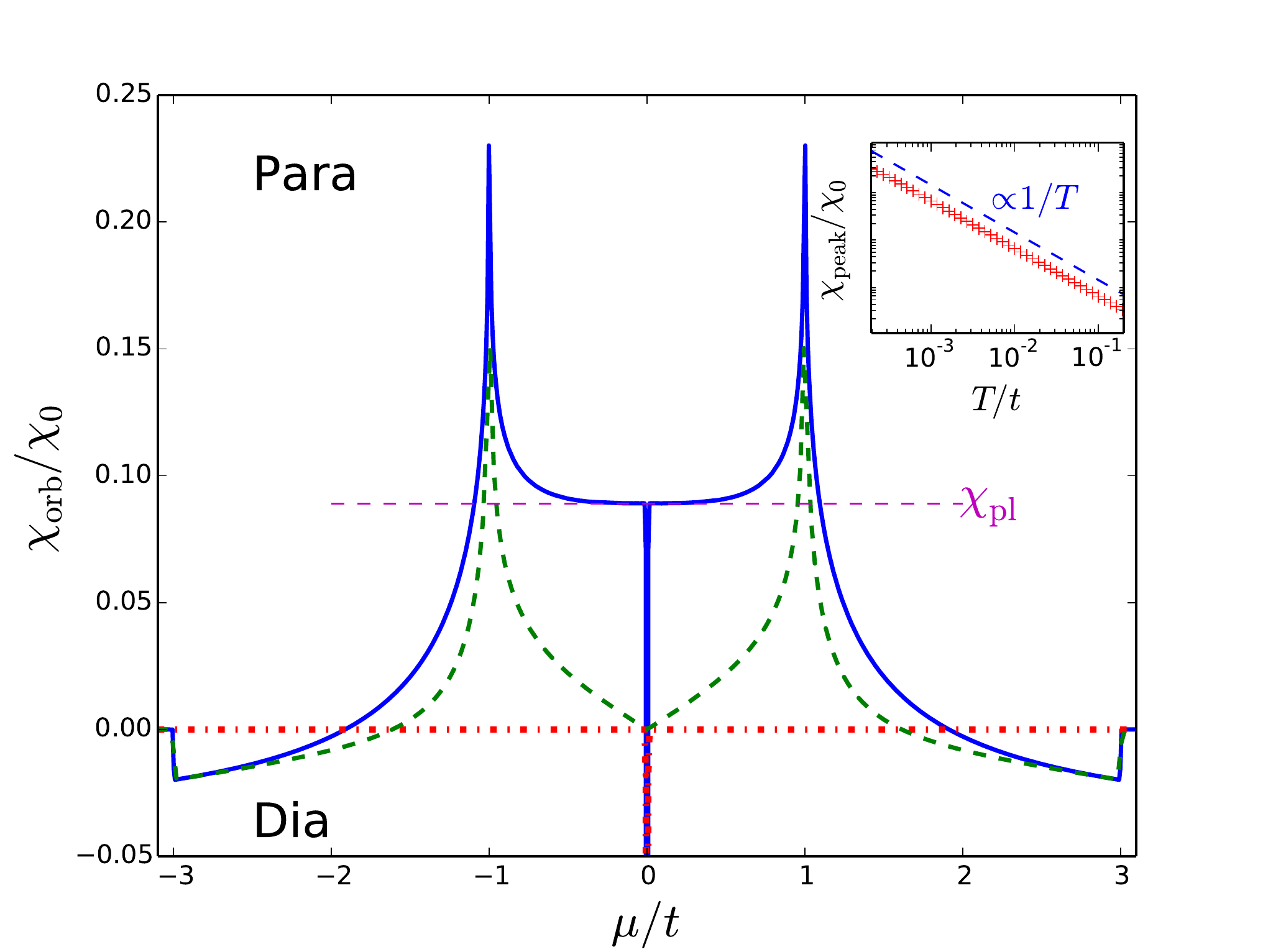}
\caption{(Color online). Chemical potential $\mu$ dependence of the orbital susceptibility $\chi_\mathrm{orb}$ at fixed temperature $T=8\cdot10^{-4}t$ for graphene. $\chi_\mathrm{orb}$ and $\mu$ are respectively in units of $\chi_0$ and $t$. Continuous line: total susceptibility; dashed line: the LP contribution; dash-dotted line: McClure's contribution. The paramagnetic plateau is also indicated as $\chi_\textrm{pl}$. Inset: $\chi_\mathrm{peak}\equiv \chi_\mathrm{orb}(\mu=0,T)-\chi_\mathrm{pl}$ as a function of temperature $T$.}
\label{fig:graphene}
\end{center}
\end{figure}
%

The total orbital susceptibility of graphene has already been derived by Gomez-Santos \emph{et al.}\cite{Gomez-Santos11} but it deserves more discussion. Fig.~\ref{fig:graphene} shows the susceptibility as a function of the chemical potential from Eq.~(\ref{eq:chi_deuxbandes}). It coincides with Fig.~1 of Ref.~[\onlinecite{Gomez-Santos11}]. The LP contribution and the McClure prediction are plotted as well. As found by McClure\cite{McClure56}, the susceptibility diverges as $-1/T$  at vanishing chemical potential:
\begin{equation}
\label{eq:chi_McClure}
\chi_\mathrm{McClure}(\mu,T)=\frac{3\chi_0}{4\pi}n_\mathrm{F}'(0)\stackrel{\mu=0}{\propto} -\frac{1}{T}.
\end{equation}
The inset of Fig.~\ref{fig:graphene} shows the result of our calculations at $\mu=0$. The qualitative behavior in $1/T$ is verified, and it quantitatively agrees with McClure up to a constant paramagnetic correction $\chi_\mathrm{pl}$ independent of $T$ which we call the \emph{paramagnetic plateau}. We found a lengthy analytical expression for this quantity in terms of an integral. Numerical quadrature gives 
\begin{equation}
\chi_\mathrm{pl}\approx0.089\chi_0\, ,\nn
\end{equation}
which is roughly five times the diamagnetism at the band edges. The numerical results at $\mu=0$ are well adjusted by $\chi_\textrm{pl}+\frac{3\chi_0}{4\pi}n_F'(0)$ with $n_F'(0)=-\frac{1}{4T}$. We compared the result for $\chi_\textrm{orb}(\mu,T)$ to our numerical method based on the Hofstadter spectrum~\cite{Raoux14} and found excellent quantitative agreement. Note that the two methods are completely different: one is a perturbative response formula, the other is non-perturbative and relies on an exact diagonalization of the Hamiltonian in a small but finite magnetic field. 

When $\mu\not=0$ at $T=0$, McClure predicted a vanishing susceptibility as Eq.~(\ref{eq:chi_McClure}) gives $\chi_\mathrm{McClure}(\mu,T)\propto-\delta(\mu)$ when $T\to0$. It appears that the strong diamagnetic contribution  is only at $\mu=0$, but we see on Fig.~\ref{fig:graphene} that the paramagnetic \emph{plateau} remains when $\mu\not=0$. 

At $\mu=\pm t$, we observe a diverging paramagnetic contribution, which corresponds to van Hove singularities in the DoS. Such orbital paramagnetism was discussed quite generally by Vignale\cite{Vignale91}. The LP formula predicts this effect, as it is encoded in the DoS and the curvature of the spectrum (see Appendix~\ref{appendix:LP_square}). 

Finally, outside of the van Hove singularities ($|\mu|\geq t$), the susceptibility is qualitatively described by the LP formula and converges in the edges of the spectrum ($\mu=\pm3t$) to the Landau diamagnetism of ``quasi-free'' electrons 
\begin{equation}
\chi_\mathrm{Landau}=-\frac{\mu_0e^2}{24\pi m^\star}=-\frac{\chi_0}{16\pi}\approx-0.02\chi_0
\end{equation}
with a band mass $m^\star=2\hbar^2/(3ta^2)$ consistent with the quadratic approximation of the dispersion relation near $\mu=\pm3t$. The diamagnetic peak and the paramagnetic \emph{plateau} should be understood as the result of strong interband coupling between valence and conduction bands. This interband coupling decreases when $|\mu|$ increases, as revealed by a better agreement with the LP formula upon approaching the band edges. Note that the sumrule (Eq.~(\ref{eq:sumrule})) is neither verified by the LP susceptibility, nor by the McClure formula alone, but is fulfilled by the total formula of Eq.~(\ref{eq:chi}).

If one is only interested in the $\mu=0$ susceptibility, one might be tempted to use a low-energy approximation. As a matter of fact, McClure partially succeeded in using this approach for graphene. With the low-energy Hamiltonian, he derived the associated Landau levels, and deduced the susceptibility by the Euler-MacLaurin formula. In a previous work\cite{Raoux14}, it has been shown that the sign of the susceptibility in a magnetic field is generally governed by the behavior of the first Landau level. Another low-energy technique, which will be used for comparison purposes in the following, is to compute the susceptibility using Eq.~(\ref{eq:chi_deuxbandes}) and a linearised Hamiltonian near zero energy (namely the Dirac points for graphene).  Appendix~\ref{appendix:linearized_graphene} gives some details for the case of graphene and recovers Eq.~(\ref{eq:chi_McClure}). The paramagnetic \emph{plateau} is however not recovered by this method; it is a property coming from the full Hamiltonian, and which can not be found by a low-energy approach. A second-order approximation does not yield the \emph{plateau} either. The \emph{plateau} contribution comes from terms proportional to $n_\mathrm{F}(\varepsilon_{s\bm k})$ in Eq.~(\ref{eq:chi_deuxbandes}). This result illustrates that the magnetic response can not be fully understood as a Fermi surface property (see, for example, the discussion in Ref.~[\onlinecite{Haldane04})].

\subsubsection{Pseudo graphene bilayer}

Bilayer graphene also has a gapless structure but with parabolic (instead of linear) band touching points. Although to be correctly described, bilayer graphene requires a $4\times 4$ Hamiltonian, the physics near the contact points is well-understood within a low-energy $2\times 2$ model\cite{McCann06}. We take a different route and choose a convenient tight-binding two-band toy-model\cite{Montambaux12}, that reproduces the low-energy effective Hamiltonian of bilayer graphene near the band touchings. As a tight-binding model, it is compatible with the use of Eq.~(\ref{eq:chi_deuxbandes}). Compared to Fig.~\ref{fig:schema}, this toy-model takes into account nearest-neighbor hopping $t$ (as in graphene) and third-nearest-neighbors with a hopping amplitude $t_3=t/2$. The Hamiltonian $h_{\bm k}$ is given by $h_{\bm k}=\bm f_\textrm{bi}\cdot\bm \sigma$ with $\bm f_\textrm{bi}=\bm f_\mathrm{gr}+\bm f_3$ and
\begin{equation}
\label{eq:f_gr13}
\bm f_3=t_3\left(
\begin{array}{c}
\cos(2k_y)+2\cos(k_y)\cos(\sqrt3k_x) \\
\sin(2k_y)-2\sin(k_y)\cos(\sqrt3k_x) \\
0
\end{array}
\right).
\end{equation}

\begin{figure}[t]
\begin{center}
\includegraphics[scale=0.4]{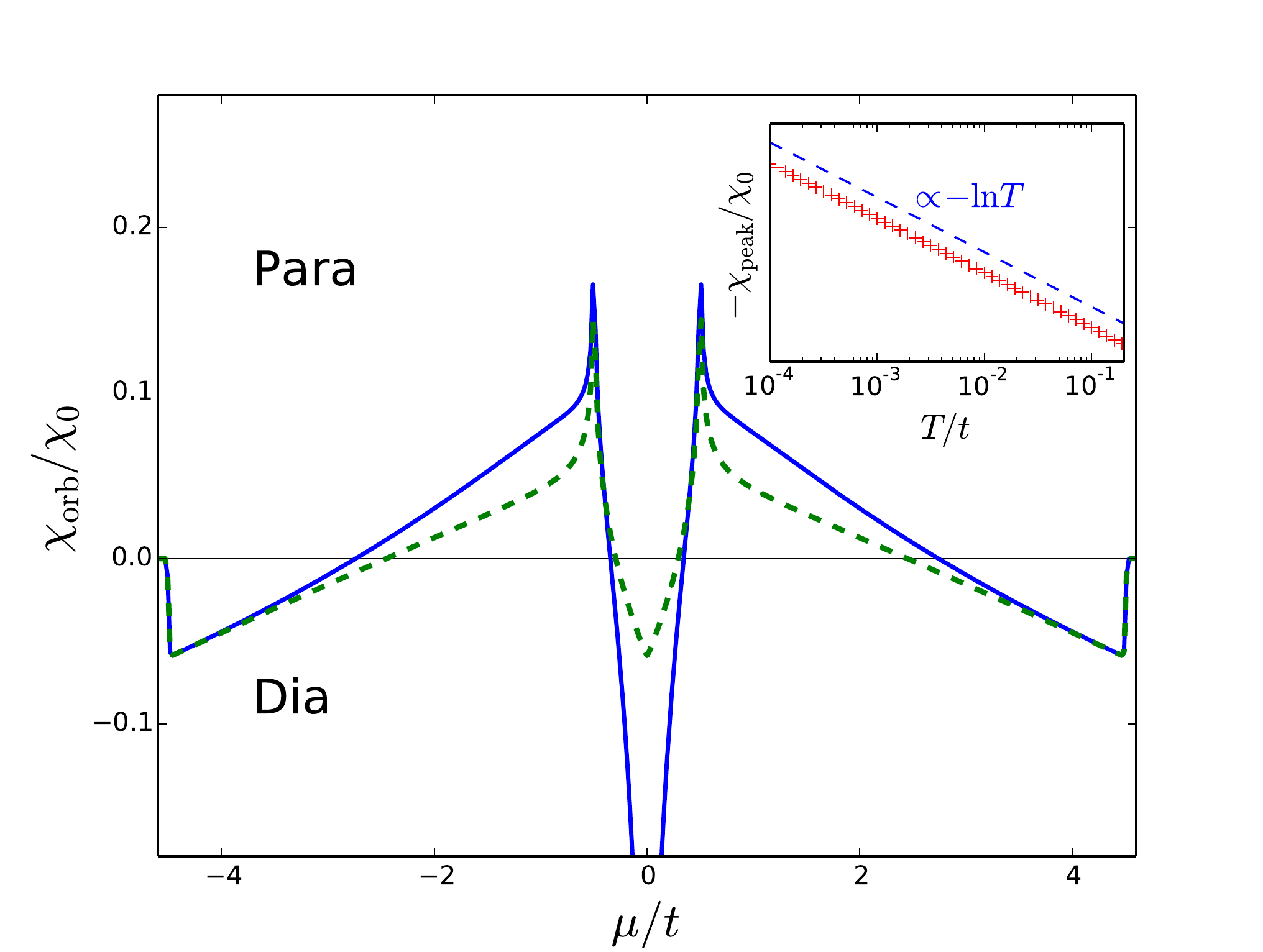}
\caption{(Color online). Chemical potential $\mu$ dependence of the orbital susceptibility $\chi_\mathrm{orb}$ at fixed temperature $T=5\cdot10^{-3}t$ for a pseudo graphene bilayer. $\chi_\mathrm{orb}$ and $\mu$ are respectively in units of $\chi_0$ and $t$. Continuous line: total susceptibility; dashed line: the LP contribution
. Inset: $\chi_\mathrm{peak}\equiv \chi_\mathrm{orb}(\mu=0,T)$ as a function of temperature $T$ in a semilog plot.}
\label{fig:t1-t3}
\end{center}
\end{figure}

Fig.~\ref{fig:t1-t3} presents the susceptibility of this model, which fulfils the susceptibility sumrule. The $\mu=0$ behavior should be similar to that of bilayer graphene (note, however, that this is not completely obvious as one of the main message of the present article is that orbital magnetism is not just a property of the Fermi surface but receives contributions from all the filled bands): there is a diamagnetic peak but with a logarithmic temperature scaling (inset of Fig.~\ref{fig:t1-t3}), different from the monolayer. The LP formula gives a finite contribution (of the Landau diamagnetism type) due to the parabolic behavior of the bands at zero energy.

A low-energy approach using the approximate Hamiltonian of bilayer graphene near $E=0$ yields
\begin{equation}
\chi_\mathrm{orb}(\mu,T)=-\frac{\chi_0}{16\pi}\int_{-t_\perp}^{t_\perp}n'_\mathrm{F}(E)\left(\ln\frac{|E|}{t_\perp}+\frac13\right)\ud E
\end{equation}
consistent with previous calculations\cite{Safran84,Koshino07}, where $t_\perp$ is an ultraviolet energy cutoff of the order of magnitude of the interlayer coupling. It gives the validity limit of the 2-band approximation of the 4-band Hamiltonian. In particular, we find that $\chi_\mathrm{orb}(\mu=0,T)\propto\ln T$. The inset of Fig.~\ref{fig:t1-t3} confirms this scaling. 


%

\subsubsection{At the merging transition: semi-Dirac fermions} 
\begin{figure}[t]
\begin{center}
\includegraphics[scale=0.4]{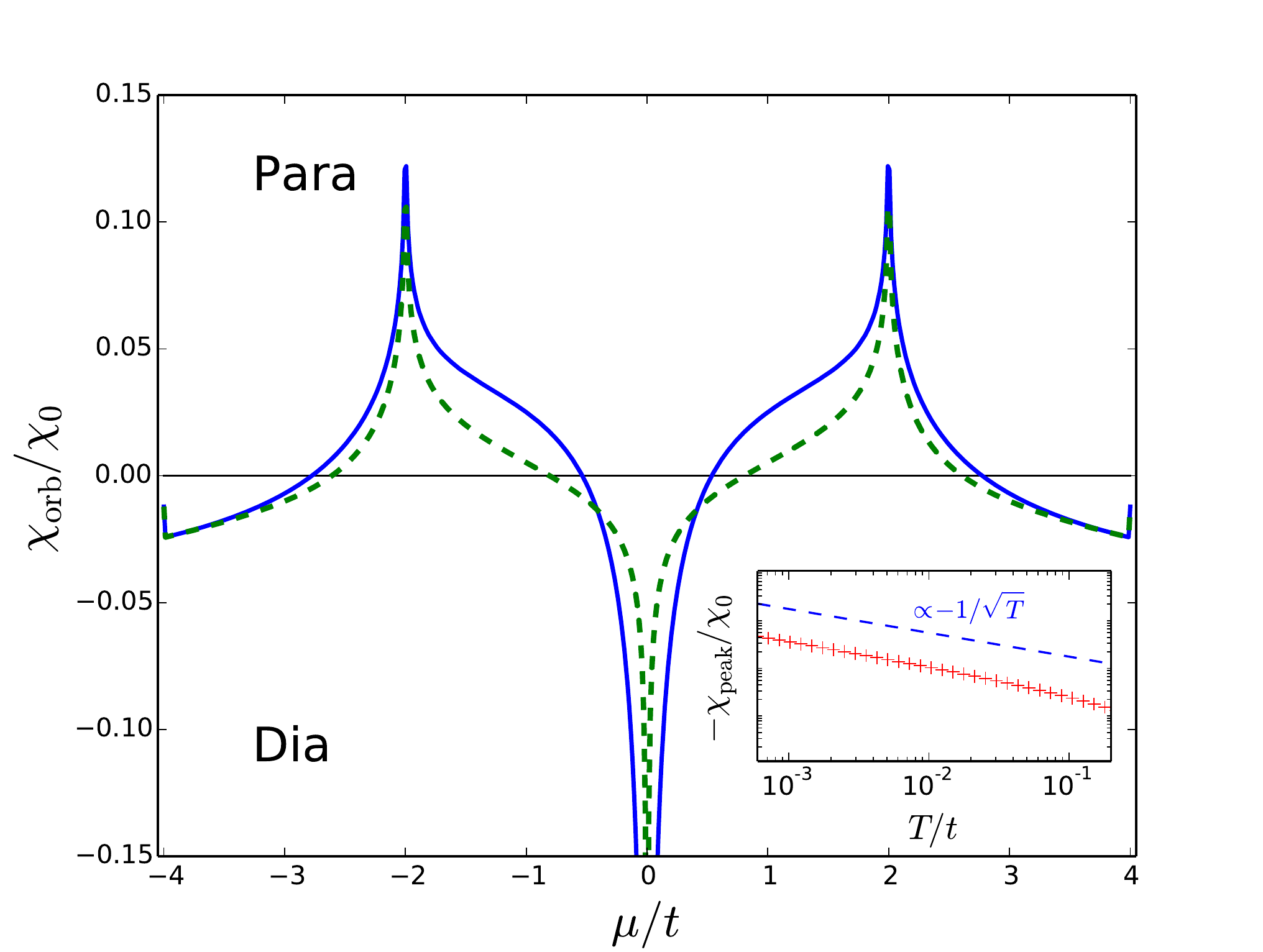}
\caption{(Color online). Chemical potential $\mu$ dependence of the orbital susceptibility $\chi_\mathrm{orb}$ at fixed temperature $T=8\cdot10^{-4}t$ for semi-Dirac fermions. $\chi_\mathrm{orb}$ and $\mu$ are respectively in units of $\chi_0$ and $t$.  Continuous line: total susceptibility; dashed line: the LP contribution. Inset: $\chi_\mathrm{peak}\equiv \chi_\mathrm{orb}(\mu=0,T)$ as a function of temperature $T$ in log-log plot.}
\label{fig:merging}
\end{center}
\end{figure}

The last gapless system we investigate is an example of semi-Dirac electrons (quadratic-linear spectrum) in a strongly deformed \mbox{honey}comb lattice described by the nearest-neighbor tight-binding model. If a uniaxial strain is applied to a graphene sheet (this is known as a quinoid deformation), it results in an anisotropy which phenomenologically induces two different values for the hopping amplitudes $t$ and $t'$\cite{Montambaux09,Montambaux09_prb} (see Fig.~\ref{fig:schema}). There is a critical point at $t'=2t$ case, where the two initial Dirac points exactly merge at a $M$ point of the Brillouin zone. This corresponds to a topological Lifshitz transition. Exactly at the transition, $h_{\bm k}=\bm f_\textrm{merg}\cdot\bm\sigma$ with\cite{Dietl08}
\begin{equation}
\label{eq:f_merging}
\bm f_\textrm{merg}=\left(
\begin{array}{c}
t'\cos(k_y)+2t\cos(k_y/2)\cos(\sqrt3k_x/2) \\
-t'\sin(k_y)+2t\sin(k_y/2)\cos(\sqrt3k_x/2) \\
0
\end{array}
\right).
\end{equation}
The corresponding low-energy spectrum (near the $M$ points) of this model is of the semi-Dirac type: it is quadratic in one reciprocal space direction (taken as $k_x$) and linear in the perpendicular ($k_y$) one. Surprinsingly, the LP formula predicts a $1/T$ diverging diamagnetic contribution at $\mu=0$, eventhough the DoS vanishes. The exact orbital susceptibility is plotted in Fig.~\ref{fig:merging}. It does show a diamagnetic divergence at zero chemical potential, but with a different temperature scaling than that predicted by the LP formula.

At $\mu=0$, a low-energy analysis can be performed analytically to yield:
\begin{equation}
\chi_\mathrm{low}(T)\approx-\frac{\chi_0}{\sqrt T} \frac{\Gamma(3/4)^2}{\pi^{5/2}}\int_0^\infty \textrm{sech}^2 x^2 \ud x
\end{equation}
%
with $\Gamma(x)$ the Euler function and $\int_0^\infty \textrm{sech}^2 x^2 \ud x \approx 0.952781$. It confirms the diamagnetic divergence and the $1/\sqrt{T}$ scaling. A similar behavior has been proposed previously\cite{Banerjee12} -- albeit with a very different numerical prefactor -- and based on an approximate one-band formula. This model gives a different example of the effect of interband coupling. In that case, it renormalizes the diverging susceptibility near the band touching. Note that, on a qualitative level, the LP formula does a reasonable job there.
%

\subsubsection{Conclusion on gapless systems}
We summarize the general behavior observed in all two-band gapless systems considered here: they all present a diamagnetic diverging susceptibility when the chemical potential is at the band touching energy. Note that this is not a general feature: in Ref.~[\onlinecite{Raoux14}], we showed that in a 3-band gapless tight-binding model, the diverging susceptibility at the band touching points could be tuned from dia- to para-magnetic, without changing the zero-field energy spectrum. The tuning parameter only affects the zero-field eigenstates. This shows the importance of interband effects.

\subsection{Gapped Systems}


\subsubsection{Boron nitride or ``gapped graphene''} 
\begin{figure}[t]
\begin{center}
\includegraphics[scale=0.4]{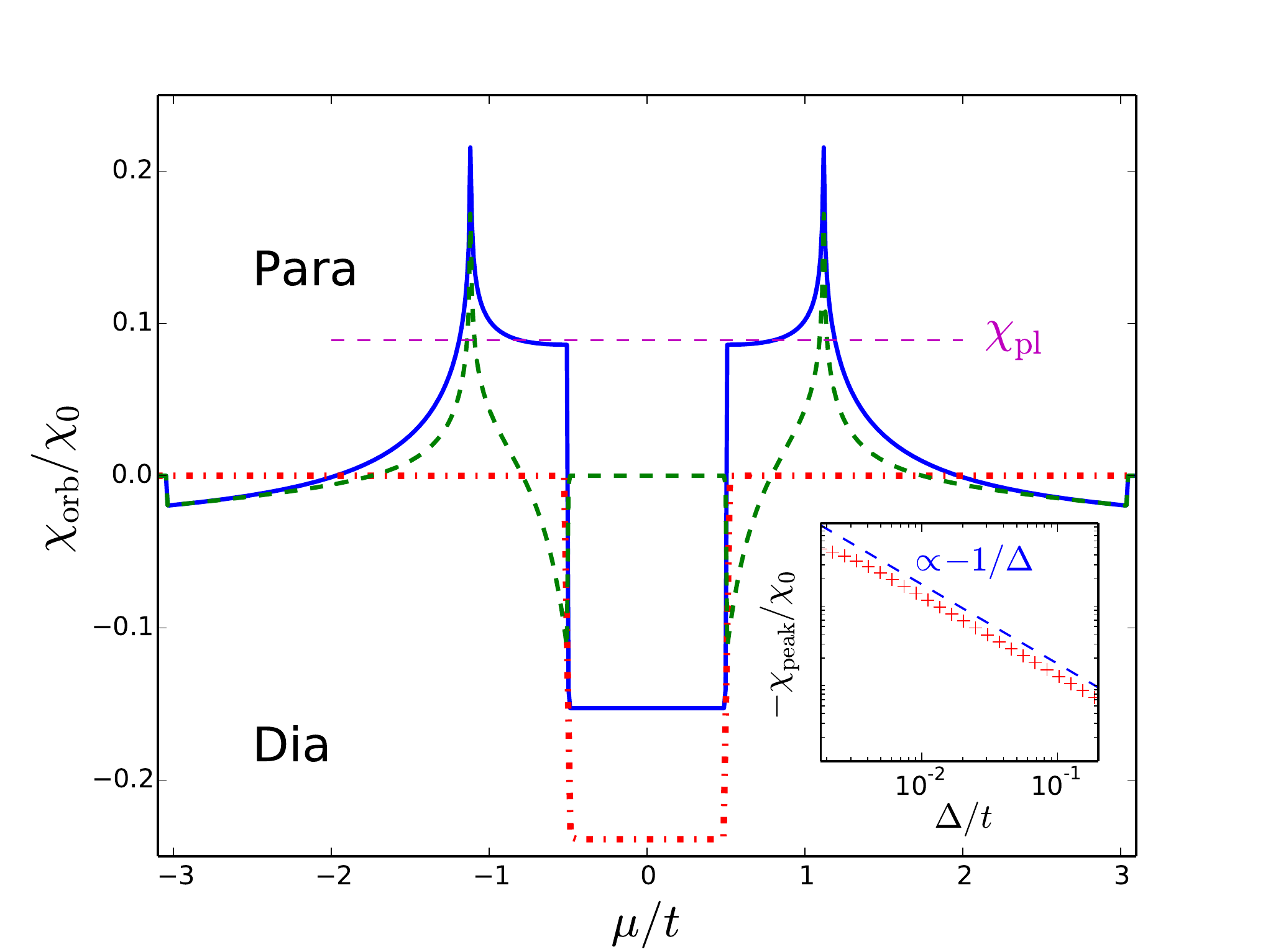}
\caption{(Color online). Chemical potential $\mu$ dependence of the orbital susceptibility $\chi_\mathrm{orb}$ at fixed temperature $T=8\cdot10^{-4}t$ for boron nitride with a gap $2\Delta=t$. $\chi_\mathrm{orb}$ and $\mu$ are respectively in units of $\chi_0$ and $t$.  Continuous line: total susceptibility; dashed line: the LP contribution. Inset: $\chi_\mathrm{peak}\equiv \chi_\mathrm{orb}(\mu=0,T)-\chi_\mathrm{pl}$ as a function of $\Delta$ in a log-log plot at fixed temperature.}
\label{fig:bn}
\end{center}
\end{figure}

Consider a honeycomb lattice with a staggered on-site potential $\pm \Delta$ for $A$ and $B$ atoms corresponding to boron and nitrogen atoms, for example. This inversion symmetry-breaking opens a gap $2\Delta$ at the \mbox{(ex-)Dirac} points\cite{Semenoff84}. The Hamiltonian is the same as that of graphene albeit with $f_\mathrm{gr}^z=\Delta$ instead of~$0$ in Eq.~(\ref{eq:f_graphene}).
In the multiband LP formula, a full or empty band does not contribute to the total magnetic response, and thus a zero susceptibility is expected in any gap. Fig.~\ref{fig:bn} shows both the LP susceptibility and the result of Eq.~(\ref{eq:chi_deuxbandes}) for boron nitride. A striking difference between the two graphs is the presence of a residual diamagnetism in the gap, even if the gap is large. Again, we interpret this observation as interband coupling even if the two bands are far away in energy. This result can be understood from the graphene case: the broadening of the $\mu=0$ peak $-\delta(\mu)$ by temperature is studied in the previous section; here the delta peak is broadened by the presence of another energy scale: namely the gap $2\Delta$. Thus, one can guess that the susceptibility in the gap depends on the gap as $1/\Delta$. This is indeed verified by the inset of Fig.~\ref{fig:bn}. More precisely, the $\mu=0$ value is the sum of two terms: the constant paramagnetic \emph{plateau} for graphene $\chi_\textrm{pl}$ and a McClure-like diamagnetic part. Using the linearised Hamiltonian of boron nitride in Eq.~(\ref{eq:chi_deuxbandes}), the low-energy approach gives:
\begin{equation}
\label{eq:chi_McClure_bn}
\chi_\mathrm{low}(\mu=0,T)=\frac{3\chi_0}{4\pi}\frac{n_\mathrm{F}(\Delta)-n_\mathrm{F}(-\Delta)}{2\Delta},
\end{equation}
which looks like a ``generalised McClure formula'' for a non-vanishing gap\cite{Nakamura07,Koshino10}. It converges towards Eq.~(\ref{eq:chi_McClure}) when $\Delta\to0$.

While crossing the gap, the susceptibility suffers a discontinuity: it goes from dia- to paramagnetism. This is surprising: at a parabolic band edge, one would expect the susceptibility to converge to Landau's diamagnetic value. The susceptibility outside of the gap is actually very similar to graphene's. A null susceptibility outside of the gap\cite{Nakamura07,Koshino10} is an artefact of the low-energy approach. When considered as a doped semiconductor, boron nitride has a very unusual behavior: near a parabolic band bottom ($|\mu|\gtrsim \Delta$), it features orbital paramagnetism $\sim \chi_\textrm{pl}\approx 0.089 \chi_0>0$, very different from the naive expectation of a Landau diamagnetism with a band mass $\chi_L=-\frac{\chi_0}{24\pi}\frac{9}{4\Delta}\sim -\frac{\chi_0}{\Delta}<0$ (both the sign and the scaling of the susceptibility with the gap are different). 
%

\subsubsection{Uniaxially strained graphene beyond the merging transition}
\begin{figure}[t]
\begin{center}
\includegraphics[scale=0.4]{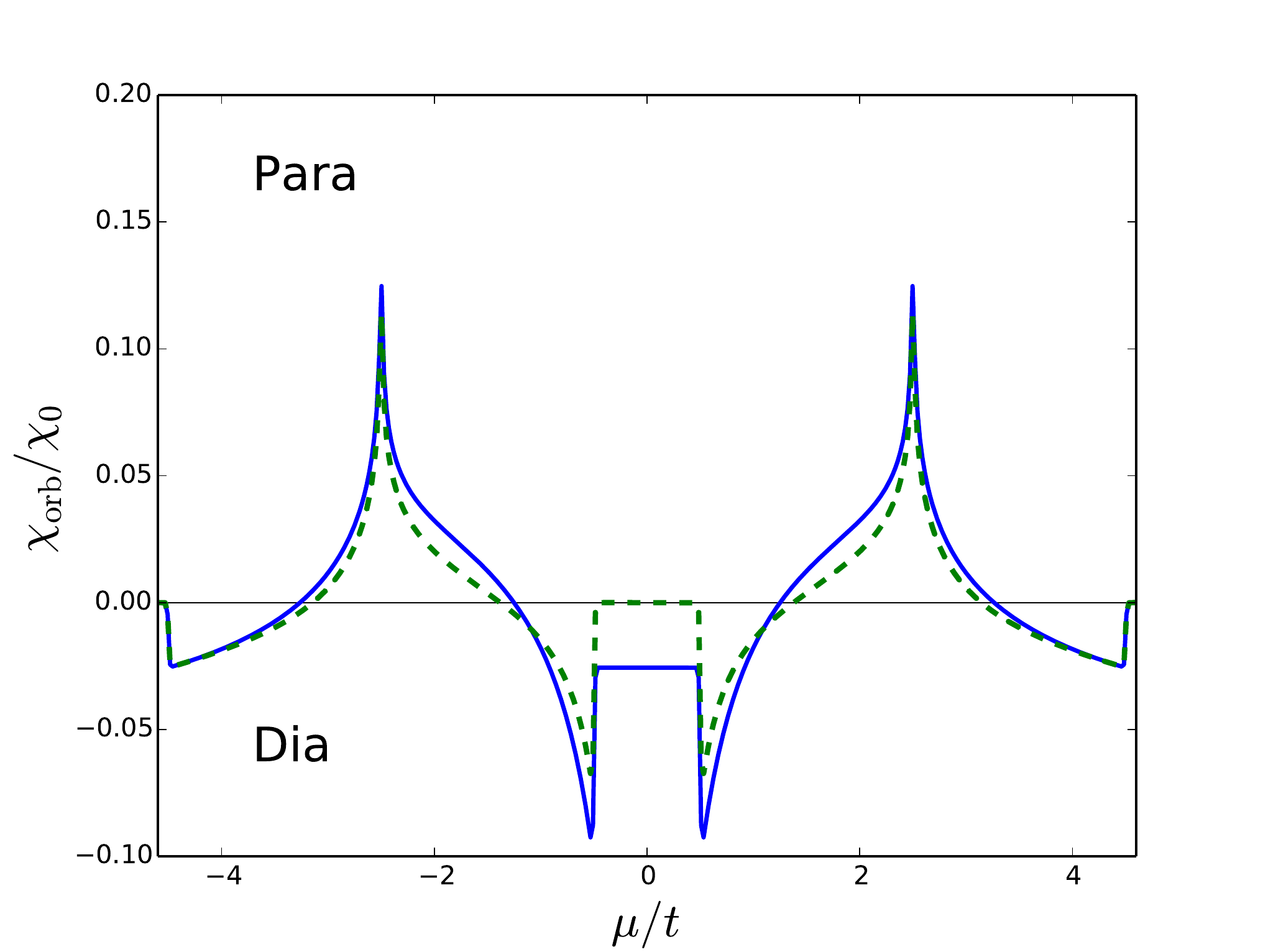}
\caption{(Color online). Chemical potential $\mu$ dependence of the orbital susceptibility $\chi_\mathrm{orb}$ at temperature $T=5\cdot 10^{-3}t$ for a uniaxially strained graphene beyond the merging transition with $t'=2.5t$ (corresponding to a gap $2\Delta_*=t$). $\chi_\mathrm{orb}$ and $\mu$ are respectively in units of $\chi_0$ and $t$.    Continuous line: total susceptibility; dashed line: the LP contribution.}
\label{fig:merging_gapped}
\end{center}
\end{figure}

Starting from the unixially strained honeycomb lattice at the merging transition (third gapless system that we studied above) and increasing the fraction $t'/t$ to a value larger than 2 (we choose $t'=2.5t$), the Dirac cones no longer exist and a gap is present in the band structure\cite{Montambaux09,Montambaux09_prb}. This case is different from boron nitride as the Dirac points have not been gapped but have completely disappeared at the merging transition. The orbital susceptibility of this model is presented in Fig.~\ref{fig:merging_gapped}. Like boron nitride, the susceptibility is diamagnetic in the band gap. However, in the vicinity of the gap, the behavior is different: it stays diamagnetic outside of the gap. Moreover, it qualitatively follows the LP formula: the merging of the Dirac points has suppressed most of the interband effects. In such a case, the Berry curvature is almost zero all over the Brillouin zone and the two valleys have disappeared at the merging transition.

A low-energy approach shows that the scaling of the susceptibility in the gap is different from boron nitride
{\color{red}}
\begin{equation}
\chi_\textrm{low}(\mu\in\textrm{gap},T=0)\propto-\frac{1}{\sqrt{\Delta_*}}
\end{equation}
where the gap is here given by $2\Delta_*=2(t'-2t)$.

%
%

\subsubsection{Conclusion on gapped systems}
The study of gapped systems showed two surprising results. First, a band insulator with a chemical potential lying inside the band gap can have a non-vanishing magnetic susceptibility\cite{Fukuyama70}. As the two models we presented above gave a constant diamagnetic susceptibility in the gap, one may argue that it is always diamagnatic and may be understood in a similar way as the diamagnetism of core electrons. However, the calculations of this paper only concern itinerant electrons (core electrons are not included). Furthermore, the study of yet another system, namely a gapped version of the $\alpha-\mathcal T_3$ lattice presented in Ref.~[\onlinecite{Raoux14}], gives a finite susceptibility in the gap that is continuously tunable from dia- to paramagnetic (without changing the zero-field spectrum).

The second surprising result obtained on gapped system is the behavior near the gap. On the gap edges, the susceptibility is by no mean forced to converge towards the Landau diamagnetic value even if the band spectrum is parabolic. Again, we interpret this result as a proof of interband coupling even for distant bands.

\section{Conclusion}
\label{sec:conclusion}

The orbital magnetism of isolated atoms was understood long ago. However that of itinerant electrons in crystalline solids has remained in an unsatisfactory state. Roughly speaking, it stayed at the basic understanding that the orbital susceptibility $\chi_\textrm{orb}$ of Bloch electrons is essentially that of free electrons (as understood by Landau\cite{Landau30}) albeit with an effective band mass $m^*$ (expected to take the band structure into account and as understood by Peierls\cite{Peierls33}): $\chi_\textrm{orb}\approx -\mu_0 \frac{e^2}{24\pi m^*}<0$ (see e.g. Ref.~[\onlinecite{Ashcroft,Abrikosov72}]). Actually, the work of Peierls contains more than that but it neglects a crucial ingredient: band coupling or interband effects. Fukuyama made an essential step by providing a compact formula including interband effects\cite{Fukuyama71}.

In the present paper, we clarify some aspects of the orbital magnetism of band coupled systems. Using a gauge-independent perturbation theory approach to compute magnetic field derivatives of the grand potential for multi-band tight-binding models, we obtained a new formula -- see Eq.~(\ref{eq:chi}) -- for the orbital susceptibility and recover the known result for the orbital magnetization -- see Eq.~(\ref{eq:M}). Then, we obtained a convenient formula for the orbital susceptibility of particle-hole symmetric two-band models -- see Eq.~(\ref{eq:chi_deuxbandes}) -- and applied it to several specific models. Equations (\ref{eq:chi}) and (\ref{eq:chi_deuxbandes}) are our main results. Here we summarize the lessons that we learned about the orbital magnetism of itinerant electrons in band coupled systems:
 \begin{itemize}
 \item According to  ``the Ashcroft and Mermin''\cite{Ashcroft,textbooks}: ``If the electrons move in a periodic potential [...], the analysis becomes quite complicated, but again results in a diamagnetic susceptibility of the same order of magnitude as the paramagnetic [Pauli] susceptibility.'' We showed that this claim is not true. First, the orbital susceptibility is not always diamagnetic, as understood long ago by Vignale\cite{Vignale91} and anticipated by Peierls\cite{footnotePeierls}. Actually, for tight-binding models, we proved a sumrule\cite{Raoux14} $\int  \chi_\textrm{orb}(\mu,T) \ud \mu=0$ that implies that the orbital susceptibility has to feature both dia- and paramagnetic behaviors as a function of the chemical potential. Secondly, the orbital susceptibility is in general not of the same order of magnitude as the Pauli susceptibility $\chi_\textrm{spin}=\mu_B^2 \rho(\mu)$ (where $\mu_B$ is the Bohr magneton). For example, the orbital susceptibility in graphene diverges at half filling as $\chi_\textrm{orb}(\mu=0,T)\approx \chi_\textrm{pl}-\frac{3\chi_0}{16\pi T}$, while the Pauli contribution goes to zero with the DoS $\rho(\mu)\propto |\mu|\to 0$.
 
\item The orbital susceptibility obtained from second order perturbation theory is not only given by the zero-field energy spectrum but crucially depends on zero-field eigenstates. Then the LP formula, which only depends on the zero-field energy spectrum, is not exact in the case of several bands. The presence of multiple bands yields interband coupling that drastically affect the susceptibility. Roughly, speaking, these eigenstates effects are thought of as geometrical properties of the Bloch bundle (collectively known as Berry phase effects in solid state physics \cite{Xiao10}) and are a measure of band coupling.

\item The Fukuyama formula\cite{Fukuyama71} does not generally apply to tight-binding models\cite{Gomez-Santos11,Koshino07}. For example, in the case of a single band, it works for the square but not for the triangular lattice tight-binding models (see Appendix \ref{appendix:LP_square}). Actually it fails for non-separable tight-binding models (in Eq.~(\ref{eq:chi_Stauber}), the Fukuyama formula corresponds to $h^{xy}=0$). As such, it does not recover the LP formula in the case of an arbitrary single band model (see the corresponding discussion in Ref.~[\onlinecite{Fukuyama71}]). Generally speaking, the Fukuyama formula is not suited for tight-binding models with a finite number of bands, as it was obtained from a different theoretical basis relying on the complete band structure made of an infinite number of bands. It could well be -- but we did not prove it -- that the Fukuyama formula follows from equation (\ref{eq:chi}) in the limit of an infinite number of bands. We suspect that the non-separability originates from the restriction to a finite number of bands of the complete Hamiltonian.

\item The orbital susceptibility is not a Fermi surface property but depends on all the filled bands (see the discussion in Ref.~[\onlinecite{Haldane04}]). It contains essential contributions from the bulk of the Fermi sea. This is best illustrated by the susceptibility at the bottom of the conduction band of boron nitride (see Fig. \ref{fig:bn}): it goes from diamagnetic in the gap to a finite paramagnetic value (roughly given by $\chi_\textrm{pl}\approx 0.089\chi_0$) as the chemical potential moves toward the bottom of the conduction band ($\mu \gtrsim \Delta$). This is in complete opposition (both in sign and magnitude) with the naive Landau diamagnetism expectation at a parabolic band edge. 
 \end{itemize}

\acknowledgements{We thank M.O. Goerbig, H. Bouchiat and collaborators for many discussions on orbital magnetism over the years.
\vspace{0.5cm}

{\it Note added:} During the completion of the present paper, Gao {\it et al.} posted a preprint on the arXiv on the geometrical effects in orbital magnetism\cite{Gao14}. Their result for the orbital susceptibility of boron nitride (see their Fig. 2a) does not agree with ours (see our Fig.~\ref{fig:bn}), which we have confirmed by exact numerical solution (Hofstadter butterfly). For example, it does not satisfy the exact sumrule. In particular, their susceptibility seems to increase near the band edges and to disagree in sign with the Landau-Peierls susceptibility, which should be diamagnetic. In addition, close to the gap ($|\mu|\gtrsim \Delta$), their susceptibility vanishes, whereas we find a paramagnetic plateau.   



\begin{thebibliography}{99}
\bibitem{footnotePeierls}Eventhough his paper is entitled ``On the diamagnetism of conduction electrons'', Peierls understood that the orbital susceptibility may change sign to become paramagnetic (see page 117 of the English translation of Ref. \onlinecite{Peierls33}).
\end{thebibliography}

\newpage

\appendix
\begin{widetext}

\section{Orbital Magnetization}
\label{appendix:magnetization}
The starting point to derive Eq.~(\ref{eq:M}), is to calculate the trace of Eq.~(\ref{eq:perturbation_1st_order}):
\begin{align}
\Tr G^{(1)}=-\frac{ieB}{2\hbar}\Tr\left(gh^xg^y- gh^yg^x\right)=-\frac{ieB}{2\hbar}\Tr\left(gh^xgh^yg- gh^ygh^xg\right)\, .
\end{align}
The Green's function can be written in terms of the Bloch state $|\psi_{n\bm k}\rangle$ and $\varepsilon_{n\bm k}$ the energy dispersion of the $n^\mathrm{th}$ band:
\begin{align}
g(E)&=\frac{S}{4\pi^2}\sum_n\int_\mathrm{BZ} \frac{\left|\psi_{n\bm k}\right\rangle\left\langle \psi_{n\bm k}\right|}{E-\varepsilon_{n\bm k}}\ud^2k\, .
\end{align}
One gets
%
\begin{equation}
\label{eq:magn}
\Tr\left\{gh^xgh^yg-gh^ygh^xg\right\}=\sum_{n,l}\frac{S}{4\pi^2}\int_\mathrm{BZ} \frac{\left\langle u_{n\bm k}\right|h_{\bm k}^x\mathcal P_{l\bm k}h_{\bm k}^y-h_{\bm k}^y\mathcal P_{l\bm k}h_{\bm k}^x\left|u_{n\bm k}\right\rangle}{(E-\varepsilon_{n\bm k})^2(E-\varepsilon_{l\bm k})}\ud^2k\, .
\end{equation}
where $h_{\bm k}=e^{-i\bm k \cdot \bm r} h e^{-i\bm k \cdot \bm r}$, $|u_{n\bm k}\rangle=e^{-i\bm k \cdot \bm r} |\psi_{n\bm k}\rangle$ is the cell-periodic part of the Bloch state and $P_{l\bm k}=|u_{l\bm k}\rangle\langle u_{l\bm k} |$ is a projector. Knowing that
\begin{equation}
\label{eq:decomposition}
\frac{1}{(E-\varepsilon_{n\bm k})^2(E-\varepsilon_{l\bm k})}=\frac{1}{\varepsilon_{n\bm k}-\varepsilon_{l\bm k}}\frac{1}{(E-\varepsilon_{n\bm k})^2}-\frac{1}{(\varepsilon_{n\bm k}-\varepsilon_{l\bm k})^2}\frac{1}{E-\varepsilon_{n\bm k}}+\frac{1}{(\varepsilon_{n\bm k}-\varepsilon_{l\bm k})^2}\frac{1}{E-\varepsilon_{l\bm k}}\, ,
\end{equation}
Eq.~(\ref{eq:magn}) can be transformed using the definitions in Eqs.~(\ref{eq:def_berry}, \ref{eq:def_orbital}) (and a change of indices for the last term of Eq.~(\ref{eq:decomposition})):
\begin{equation}
M(\mu)=\frac{e}{\hbar}\frac{T}{2\pi}\Im m\int_{-\infty}^{+\infty}\ud E \ln\left(1+e^{-\beta(E-\mu)}\right)\int_\mathrm{BZ} \frac{\ud^2k}{4\pi^2} \sum_n\left\{\frac{2\hbar}{e}\frac{m_{n\bm k}}{(E-\varepsilon_{n\bm k})^2}-2\frac{\Omega_{n\bm k}}{(E-\varepsilon_{n\bm k})}\right\}\, .
\end{equation}
Computing the integral over the energy with the formula
\begin{equation}
\label{eq:integral_E}
\Im m\int_{-\infty}^{+\infty} \frac{f(E)}{(E-\alpha)^j}\ud E=-\frac{\pi}{(j-1)!}f^{(j-1)}(\alpha)
\end{equation}
allows us to finish the calculation and to recover Eq.~(\ref{eq:M}).
%
%

\section{Partial Integration for Tr$G^{(2)}$}
\label{appendix:2nd_order}

Starting from Eq. (23), and using the identity $g^{ij}=gh^{ij}g+gh^{i}gh^{j}g+gh^{i}gh^{j}g$,
the trace of $ G^{(2)}$ reads:
\be
{\rm Tr} \lbrace G^{(2)}\rbrace = -\frac{e^2B^2}{8\hbar^2}( A_1 + 2 A_2 + 2 A_3)
\label{B1}
\ee
with
\be
\begin{array}{l}
A_1={\rm Tr} \lbrace g \left[ (h^{xx} g h^{yy} g+h^{yy} g h^{xx} g)- 2(h^{xy} g)^2 \right] \rbrace,  \\
A_2={\rm Tr} \lbrace g \left[  h^{xx} g(h^y g)^2+ h^{yy} g (h^x g)^2 - h^{xy} ( h^y g h^x g+ g  h^x g h^y g)\right] \rbrace. \\
A_3={\rm Tr} \lbrace g ([ h^x  g, h^y  g])^2    \rbrace \\
\end{array}
\ee
Using the identity $\partial_{E} g=-( g)^2$ and the cyclicity of the trace $\textrm{Tr}\left \lbrace \right\rbrace$, 
one verifies

\be
\begin{array}{l}
A_1=-\partial_{E} B_1,\\
A_2=\partial_{E} (B_2+2 B_3)  -A_3,
\end{array}
\label{B2}
\ee
where 
\be
\begin{array}{l}
B_1= {\rm Tr} \lbrace h^{xx} g h^{yy} g - (h^{xy} g)^2 \rbrace , \\
B_2= {\rm Tr} \lbrace  g h^{xy} (h^y g h^x g+ g  h^x g h^y g) \rbrace ,\\
B_3= {\rm Tr} \lbrace  (g h^x)^2 (g h^y)^2 \rbrace .
\end{array}
\ee
Using again the cyclicity of the trace, one can further establish the identity:
\be
B_1=3B_2+4B_3+2B_4,
\label{B3}
\ee
where $B_4={\rm Tr} \lbrace  (g h^x g h^y)^2 \rbrace$.
Using Eqs. (\ref{B1}, \ref{B2}, \ref{B3}), one finally obtains the following three equivalent writtings:
\be
\begin{array}{ll}
{\rm Tr} \lbrace G^{(2)}\rbrace& = \frac{e^2B^2}{8\hbar^2}\partial_{E}  (B_1-2B_2-4B_3) , \\
& = \frac{e^2B^2}{24\hbar^2}\partial_{E}  (B_1-4(B_4-B_3) ) , \\
& = \frac{e^2B^2}{8\hbar^2}\partial_{E}  (B_2+2B_4),
\end{array}
\ee
where the second line corresponds to Eq.~(23) that allows to obtain
susceptibility formula Eq.~(24), whereas the third line allows to recover susceptibility formula Eq. (26) first derived in \cite{Gomez-Santos11}.
Note that $N^{(2)}(E,B)=-\frac{e^2B^2}{24\hbar^2}\frac{\Im m}{\pi}  (B_1-4(B_4-B_3) )$ 
represents the second order correction to the integrated DoS.
Since the total number of states is magnetic field independent, $N^{(2)}(E)$ necessarily vanishes outside the zero-field energy bandwidth.

\section{Landau-Peierls formula for single-band models}
\label{appendix:LP_square}
For a single band tight-binding model, the LP susceptibility
\begin{equation}
\label{eq:square}
\chi_\mathrm{LP}(\mu,T)=\frac{\mu_0e^2}{12\hbar^2}\int_{BZ} n'_\mathrm{F}(\varepsilon_{\bm k})\left(\varepsilon_{\bm k}^{xx}\varepsilon_{\bm k}^{yy}-(\varepsilon_{\bm k}^{xy})^2\right)\frac{\ud^2k}{4\pi^2}
\end{equation}
is exact (it is easily derived from Eq. (\ref{eq:chi})). The magnetic response of the square and triangular lattices are here investigated in order to illustrate the (not so well-known) physics contained in this formula.

Note that, apart from the derivative of the Fermi function, the integrand of Eq.~(\ref{eq:square}) can be understood as the determinant of the Hessian matrix of the spectrum $\varepsilon_{\bm k}$. The susceptibility is thus governed by an intrinsic geometrical quantity of the spectrum: the Hessian $\mathcal{H}_{\bm k}=\varepsilon_{\bm k}^{xx}\varepsilon_{\bm k}^{yy}-(\varepsilon_{\bm k}^{xy})^2$, which is almost the Gaussian curvature of the band spectrum. When the spectrum can be approximate to a quadratic dispersion, it reads $\mathcal{H}_{\bm k}=\frac{\hbar^4}{m_1m_2}$ with $m_1$ and $m_2$ the two effective masses of the spectrum. Thus, for a parabolic spectrum at zero temperature
\begin{equation}
\label{eq:vignale}
\chi_\mathrm{LP}=-\frac{\mu_0e^2}{12\hbar^2}\frac{\hbar^4}{m_1m_2}\rho(\mu)=-\frac{\mu_0e^2\hbar^2}{12|m_1m_2|}\rho(\mu)\times \textrm{sgn}(m_1m_2)
\end{equation}
with $\rho(\mu)$ the DoS at the Fermi energy $\mu$. From Eq.~(\ref{eq:vignale}), we deduce that the LP susceptibility diverges when the DoS does; and $\chi_\mathrm{LP}$ can change sign at a saddle point, \emph{i.e.} when $m_1$ and $m_2$ have different signs\cite{Vignale91}. The physical picture behind such a behavior was provided by Vignale\cite{Vignale91}: it is that of a counter-circulating orbit around the saddle point. The latter is made of pieces of four regular cyclotron orbits connected by quantum mechanical tunneling events (known as magnetic breakdown in this context).

\subsection{Square lattice}
The dispersion relation for the square lattice (with $t=1$ and $a=1$) 
\begin{equation}
\varepsilon_{\bm k}=-2\cos k_x-2\cos k_y
\end{equation}
is separable. We find a simple susceptibility (with $\chi_0\equiv \mu_0 e^2 t a^2/\hbar^2=\mu_0 e^2$):
\begin{equation}
\chi_\mathrm{LP}(\mu,T)=\frac{\chi_0}{12\pi^2}\int_{BZ} n'_\mathrm{F}(\varepsilon_{\bm k})\cos k_x\cos k_y\ud k_x\ud k_y\stackrel{T=0}{=}\frac{\chi_0}{6\pi^2}Q_{1/2}\left(1-\frac{\mu^2}{8}\right)
\end{equation}
where $Q_\alpha(x)$ is the Legendre function of the second kind. This compact formula is consistent with another one proposed in Ref.~[\onlinecite{Skudlarski91}]. This is plotted in Fig.~(\ref{fig:square}). 
\begin{figure}[h]
\begin{center}
\includegraphics[width=12cm]{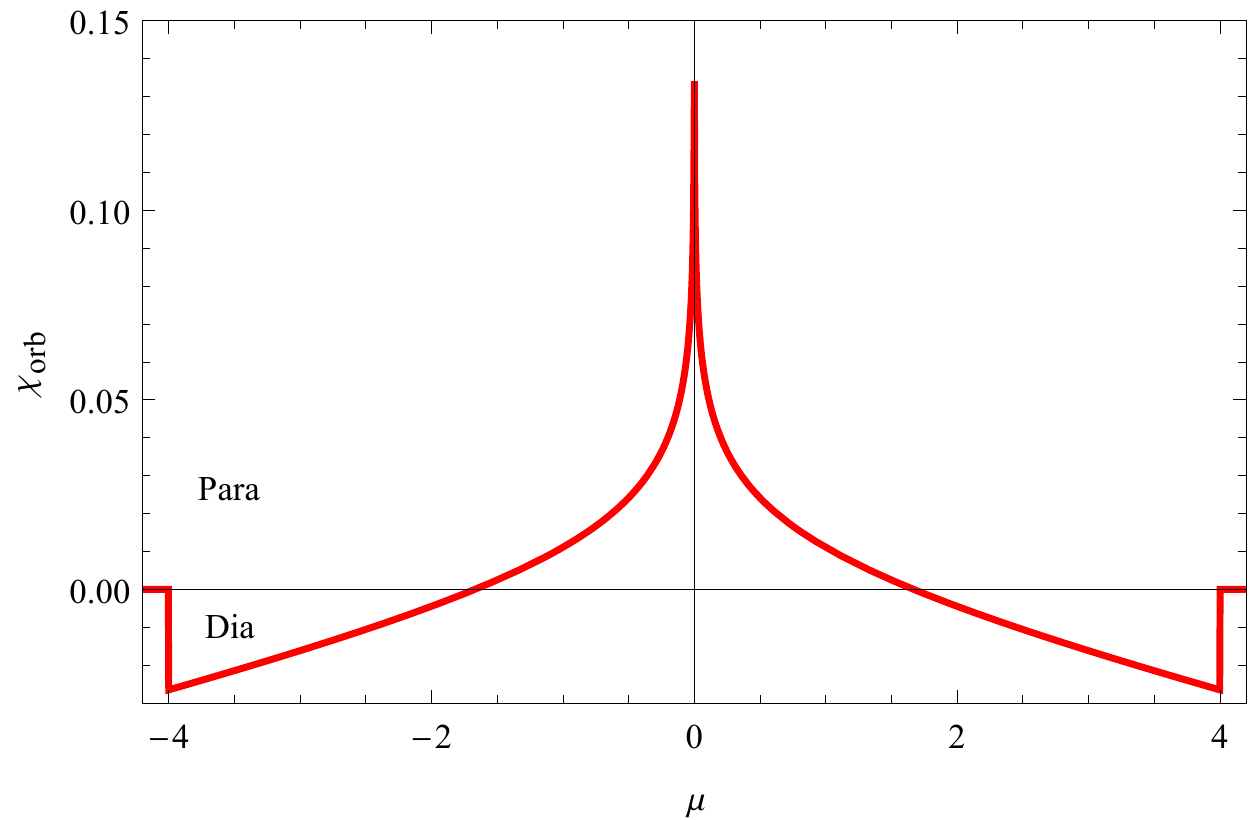}
\caption{(Color online). Orbital susceptibility $\chi_\textrm{orb}$ (in units of $\chi_0$) as a function of the chemical potential $\mu$ (in units of the hopping amplitude $t$) at $T=0$ for the tight-binding model on a square lattice.}
\label{fig:square}
\end{center}
\end{figure}

Some features of this figure are worth commenting on:
\begin{enumerate}
\item The susceptibility can be either positive or negative. The situation is different from free particles where the susceptibility is always negative (Landau diamagnetism).
\item The susceptibility verifies the sumrule: $\int\chi_\mathrm{orb}\ud \mu=0$ and thus it has both dia- or para-magnetic behavior depending on $\mu$.
\item In the vicinity of the band edges (band bottom and band top), the susceptibility tends to a diamagnetic value, which is exactly the Landau susceptibility $\chi_\mathrm{L}=-\mu_0e^2/(24\pi m^\star)$ with an effective band mass $m^\star=\hbar^2/(2ta^2)$. In the limit of low/high filling, we recover free electrons/holes with an effective mass.
\item At half filling (vanishing chemical potential), the susceptibility is paramagnetic and diverges logarithmically. This is a consequence of the van Hove singularity in the DoS. The latter is related to saddle points in the spectrum, at which the effective masses $m_1$ and $m_2$ are of different signs, leading to orbital paramagnetism.
\end{enumerate}

\subsection{Triangular lattice}
The nearest-neighbor tight-binding model on the triangular lattice is quite interesting as it has a dispersion relation which is not separable. This will allow us to compare several predictions for the orbital susceptibility. The dispersion relation is (with $t=1$ and $a=1$)
\be
\varepsilon_{\bm k}=2\cos k_x + 2 \cos\frac{k_x+\sqrt{3}k_y}{2}+2 \cos\frac{-k_x+\sqrt{3}k_y}{2}.
\ee
Since there is only one band, the exact orbital susceptibility is given by the LP formula (\ref{eq:square}) and satisfies the sumrule. Here, in contrast to the square lattice, $\varepsilon^{xy}_{\bm k}\neq 0$. The result is plotted in Fig.~\ref{fig:triangular}.
\begin{figure}[t]
\begin{center}
\includegraphics[width=12cm]{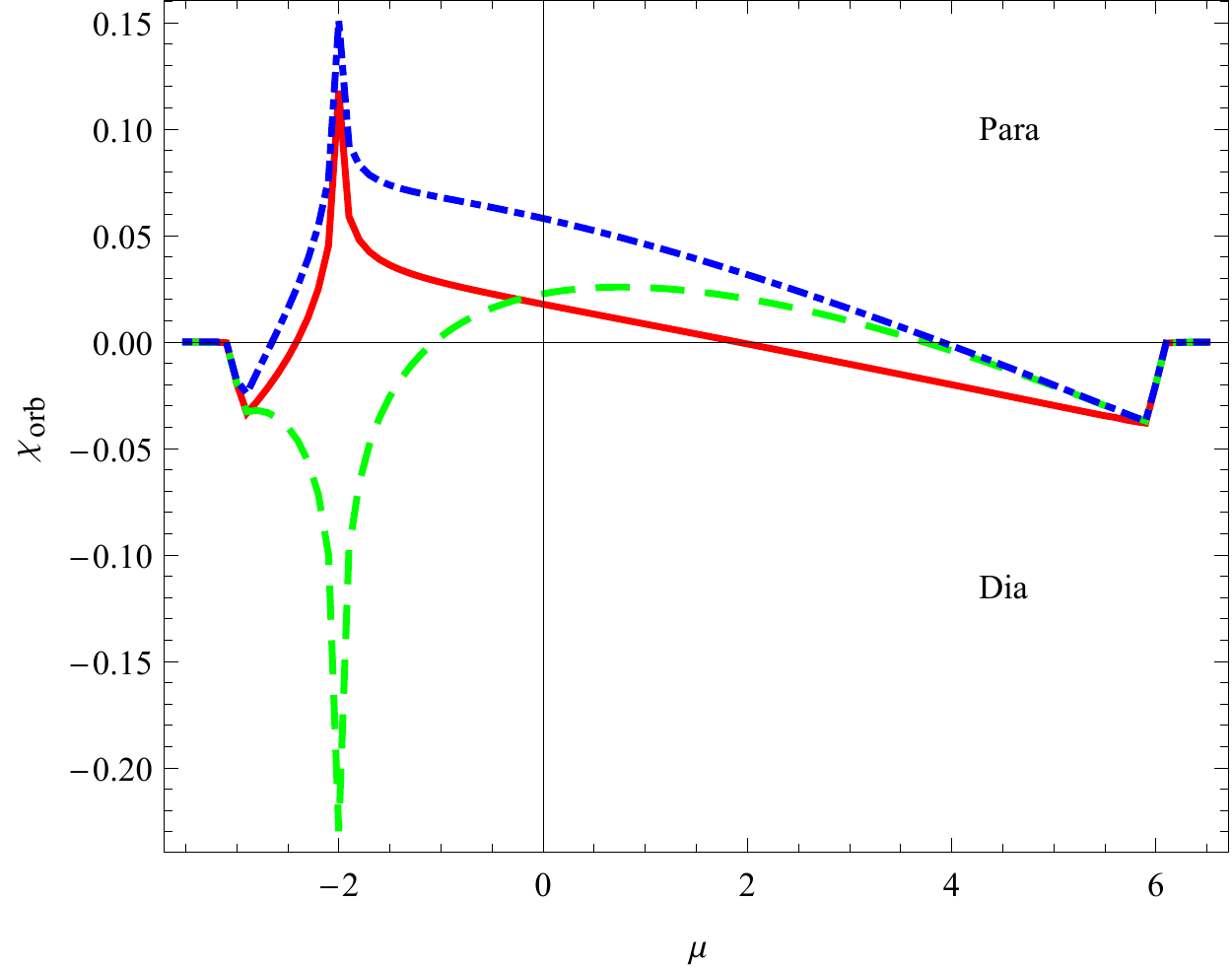}
\caption{(Color online). Orbital susceptibility $\chi_\textrm{orb}$ (in units of $\chi_0$) as a function of the chemical potential $\mu$ (in units of the hopping amplitude $t$) at $T=0$ for the tight-binding model on a triangular lattice (we took a level broadening of $\eta=5. 10^{-3}t$). LP susceptibility is shown in full line (red), Fukuyama in dashed (green) and HS in dot-dashed (blue).}
\label{fig:triangular}
\end{center}
\end{figure}

Other single-band formulas, which exist in the literature, disagree with the above exact result. For example, application of the Fukuyama formula gives the following susceptibility\cite{Fukuyama71}:
\begin{equation}
\label{eq:fuku1band}
\chi_\mathrm{F}(\mu,T)=\frac{\mu_0e^2}{12\hbar^2}\int_\mathrm{BZ} n'_\mathrm{F}(\varepsilon_{\bm k})\left(\varepsilon_{\bm k}^{xx}\varepsilon_{\bm k}^{yy}+2(\varepsilon_{\bm k}^{xy})^2+\frac{3}{2}[\varepsilon_{\bm k}^{x}\varepsilon_{\bm k}^{xyy}+\varepsilon_{\bm k}^{y}\varepsilon_{\bm k}^{yxx}]\right)\frac{\ud^2k}{4\pi^2}=\frac{\mu_0e^2}{12\hbar^2}\int_\mathrm{BZ} n'''_\mathrm{F}(\varepsilon_{\bm k})(\varepsilon_{\bm k}^{x}\varepsilon_{\bm k}^{y})^2 \frac{\ud^2k}{4\pi^2}\, .
\end{equation}
It is obtained from Eq.~(\ref{eq:chi_Stauber}) by keeping only the first term (with four Green's functions $g$) and restricting to a single band model. On the second expression above (involving $n_F'''$), it is easy to check that it satisfies the sumrule. It is also plotted in Fig.~\ref{fig:triangular}.

Another formula was proposed by Hebborn and Sondheimer (HS), which in the single band case reduces to\cite{Fukuyama71}
\begin{equation}
\label{eq:hs1band}
\chi_\mathrm{HS}(\mu,T)=\frac{\mu_0e^2}{12\hbar^2}\int_\mathrm{BZ} n'_\mathrm{F}(\varepsilon_{\bm k})\left(\varepsilon_{\bm k}^{xx}\varepsilon_{\bm k}^{yy}-(\varepsilon_{\bm k}^{xy})^2+\frac{3}{2}[\varepsilon_{\bm k}^{x}\varepsilon_{\bm k}^{xyy}+\varepsilon_{\bm k}^{y}\varepsilon_{\bm k}^{yxx}]\right)\frac{\ud^2k}{4\pi^2}
\end{equation}
It is almost identical to the expression for $\chi_F$ except that it involves the Hessian $\mathcal{H}_{\bm k}$. However, it does not satisfy the sumrule. It is plotted in Fig.~\ref{fig:triangular} as well.

\subsection{Conclusion on the orbital susceptibility in the case of a single band}
On the one-hand, the tight-binding model on the square lattice is separable ($\varepsilon_{\bm k}^{xy}=0$). In this case, the LP, the Fukuyama and the HS susceptibilities all agree. On the other hand, the tight-binding model on the triangular lattice is not separable ($\varepsilon_{\bm k}^{xy}\neq 0$) and allows one to discriminate between the different predictions. The LP susceptibility is exact in the case of a single band (whether separable or not). In addition, it does satisfy the sumrule. The Fukuyama formula also satisfies the sumrule, but it strongly disagrees with the exact result. For example, it predicts a strange diamagnetic peak at the van Hove singularity (when a paramagnetic peak is generally expected\cite{Vignale91}). The HS susceptibility is closer to the exact result but is also wrong and does not satisfy the sumrule.

\section{2-band derivation}
\label{appendix:2band}

\subsection{Definitions}
The aim of this section is to derive a computable two-band formula for the susceptibility of particle-hole symmetric systems. In order to use Eq.~(\ref{eq:chi}), we first need to compute the successive derivatives of 
%
%
$h$ (where $h$ is here a shorthand notation for $h_{\bm k}$) (as described in Sec.~\ref{sec:two-band}). They read:
\begin{align}
h&=\varepsilon\bm n\cdot\bm \sigma\\ 
h^i&=\varepsilon^i\bm n\cdot\bm \sigma +\varepsilon\bm n^i\cdot\bm \sigma\\ 
h^{ij}&=(\varepsilon^{ij}-\varepsilon\bm n^i\cdot\bm n^j)\bm n\cdot\bm \sigma+\bm a^{ij}\cdot\bm\sigma
\end{align}
with $\bm a^{ij}=\varepsilon^i\bm n^j+\varepsilon^j\bm n^i+\varepsilon\bm n\times(\bm n^{ij}\times \bm n)$.

The effect of a projector onto the derivatives of $h$ is given by:
\begin{align}
\mathcal P_sh^i\mathcal P_s&=s\varepsilon^i\mathcal P_s \label{Ps} \\
\mathcal P_sh^i\mathcal P_{-s}&=\frac{\varepsilon}{2}(\bm n^i+is\bm n\times\bm n^i)\cdot\bm\sigma \label{P-s}\\
\mathcal P_sh^{ij}\mathcal P_s&=s(\varepsilon^{ij}-\varepsilon\bm n^i\cdot\bm n^j)\mathcal P_s\\
\mathcal P_sh^{ij}\mathcal P_{-s}&=\frac12\left(\bm a^{ij}+is\bm n\times\bm a^{ij}\right)\cdot\bm\sigma.
\end{align}
Finally, the following identity will reveal to be useful in the following:
\begin{equation}
g_+g_-=\frac{1}{2\varepsilon}(g_+-g_-).
\end{equation}

\subsection{Two-Green's functions term}
$U=\tr\left\{gh^{xx}gh^{yy}-gh^{xy}gh^{xy}\right\}$ is first investigated. On the eigenprojectors basis,
\begin{align}
U&=\sum_{s} g_s^2C_{ss}+\sum_{s} g_sg_{-s}C_{-ss}
\end{align}
with 
\begin{align}
C_{ss'}=\tr\left\{\mathcal P_sh^{xx}\mathcal P_{s'}h^{yy}-\mathcal P_sh^{xy}\mathcal P_{s'}h^{xy}\right\}\, . 
\end{align}
After some algebra, one gets:
\begin{align}
&C_{ss}=(\varepsilon^{xx}-\varepsilon\bm n^x\cdot\bm n^x)(\varepsilon^{yy}-\varepsilon\bm n^y\cdot\bm n^y)-(\varepsilon^{xy}-\varepsilon\bm n^x\cdot\bm n^y)^2 \\
&C_{-ss}=\bm a^{xx}\cdot\bm a^{yy}-\bm a^{xy}\cdot\bm a^{xy}-is(\bm a^{xx}\times\bm a^{yy})\cdot\bm n
\end{align}
such that:
\begin{align}
\label{eq:U}
U&=(g_+^2+g_-^2)\left[(\varepsilon^{xx}-\varepsilon\bm n^x\cdot\bm n^x)(\varepsilon^{yy}-\varepsilon\bm n^y\cdot\bm n^y)-(\varepsilon^{xy}-\varepsilon\bm n^x\cdot\bm n^y)^2\right] +2g_+g_-(\bm a^{xx}\cdot\bm a^{yy}-\bm a^{xy}\cdot\bm a^{xy})
\end{align}
The second term of Eq.~(\ref{eq:U}) can be splited in two such that
\begin{align}
U&=(g_+^2+2g_+g_-+g_-^2)U_1+2g_+g_-U_2\\
&=4\varepsilon^2g_+^2g_-^2U_1+2g_+g_-U_2
\end{align}
where
\begin{align}
U_1&= (\varepsilon^{xx}-\varepsilon\bm n^x\cdot\bm n^x)(\varepsilon^{yy}-\varepsilon\bm n^y\cdot\bm n^y)-(\varepsilon^{xy}-\varepsilon\bm n^x\cdot\bm n^y)^2\\
U_2&= (\varepsilon\bm n)^{xx}\cdot(\varepsilon\bm n)^{yy}-(\varepsilon\bm n)^{xy}\cdot(\varepsilon\bm n)^{xy}
\end{align}

\subsection{Four-Green's functions term}
$V=\tr(gh^xgh^xgh^ygh^y-gh^xgh^ygh^xgh^y)$ will be evaluated using the same method. Similarly,
\begin{align}
V&=\sum_s \left[ g_s^4C_{ssss}+g_s^3g_{-s}(C_{-ssss}+C_{s-sss}+C_{ss-ss}+C_{sss-s})+g_s^2g_{-s}^2(C_{-s-sss}+C_{-ss-ss}+C_{-sss-s}\right]
\end{align}
where the coefficients $C_{ss's''s'''}$ are defined by
\begin{equation}
C_{ss's''s'''}=\tr\left\{\mathcal P_sh^{x}\mathcal P_{s'}h^{x}\mathcal P_{s''}h^{y}\mathcal P_{s'''}h^{y}-\mathcal P_sh^{x}\mathcal P_{s'}h^{y}\mathcal P_{s''}h^{x}\mathcal P_{s'''}h^{y}\right\}.
\end{equation}
Using Eqs.(\ref{Ps}, \ref{P-s}), in the definition of $C_{ss's''s'''}$ gives
\begin{align}
&C_{ssss}=\varepsilon^x\varepsilon^x\varepsilon^y\varepsilon^y-\varepsilon^x\varepsilon^y\varepsilon^x\varepsilon^y\
=0\\
&C_{-ssss}=0\\
&C_{s-sss}=(\varepsilon^y)^2\varepsilon^2\bm n^x\cdot\bm n^x- \varepsilon^x\varepsilon^y\varepsilon^2(\bm n^x\cdot\bm n^y-is\bm n^x\times\bm n^y\cdot \bm n)\\
&C_{ss-ss}=0\\
&C_{sss-s}=(\varepsilon^x)^2\varepsilon^2\bm n^y\cdot\bm n^y- \varepsilon^x\varepsilon^y\varepsilon^2(\bm n^x\cdot\bm n^y+is\bm n^x\times\bm n^y\cdot \bm n)\\
&C_{-s-sss}=-\varepsilon^x\varepsilon^y\varepsilon^2(\bm n^x\cdot\bm n^y-is\bm n^x\times\bm n^y\cdot \bm n)+(\varepsilon^x)^2\varepsilon^2\bm n^y\cdot\bm n^y\\
&C_{-sss-s}=-\varepsilon^x\varepsilon^y\varepsilon^2(\bm n^x\cdot\bm n^y+is\bm n^x\times\bm n^y\cdot \bm n)+(\varepsilon^y)^2\varepsilon^2\bm n^x\cdot\bm n^x\\
&C_{-ss-ss}=\frac{\varepsilon^4}{2}\left((\bm n^x\cdot\bm n^x)(\bm n^y\cdot\bm n^y)-(\bm n^x\cdot\bm n^y)^2+3((\bm n^x\times\bm n^y)\cdot\bm n)^2\right),
\end{align}
such that
\begin{align}
V&=\varepsilon^2(g_+^3g_-+g_+g_-^3+2g_+^2g_-^2)\left[(\varepsilon^y)^2\bm n^x\cdot\bm n^x+(\varepsilon^x)^2\bm n^y\cdot\bm n^y-2\varepsilon^x\varepsilon^y\bm n^x\cdot\bm n^y\right]\\
&+2g_+^2g_-^2\frac{\varepsilon^4}{2}\left[(\bm n^x\cdot\bm n^x)(\bm n^y\cdot\bm n^y)-(\bm n^x\cdot\bm n^y)^2+3((\bm n^x\times\bm n^y)\cdot\bm n)^2\right]\nonumber \\
&=\varepsilon^2(g_+^3g_-+g_+g_-^3+2g_+^2g_-^2)\left(\varepsilon^y\bm n^x-\varepsilon^x\bm n^y\right)^2 +4\varepsilon^4g_+^2g_-^2(\left(\bm n^x\times\bm n^y\right)\cdot\bm n)^2\\
&=\varepsilon^2(g_+^3g_-+g_+g_-^3+2g_+^2g_-^2)V_1 +4\varepsilon^2g_+^2g_-^2V_2.
\end{align}

\subsection{Results}

The product of Green's functions has to be decomposed to compute the integral over the energy:
\begin{align}
2g_+g_-&=\frac{1}{\varepsilon}(g_+-g_-) \\
4\varepsilon^2g_+^2g_-^2&=g_+^2+g_-^2-\frac{g_+-g_-}{\varepsilon}\\
\varepsilon^2(g_+^3g_-+g_+g_-^3+2g_+^2g_-^2)&=\frac\varepsilon2 (g_+^3-g_-^3)+\frac{1}{4}(g_+^2+g_-^2)-\frac{1}{4\varepsilon}(g_+-g_-)
\end{align}

and using 
\begin{equation}
\Im m\int_{-\infty}^{+\infty}\frac{n_\mathrm{F}(E)}{(E-\varepsilon_s)^k}\ud E=-\frac{\pi}{k!} n_\mathrm{F}^{(k)}(\varepsilon_s)
\end{equation}
where $n^{(k)}_\mathrm F$ is the $k^\textrm{th}$ derivative of $n_\mathrm F$, the integral over $E$ can be performed
\begin{align}
&\chi_\mathrm{orb}(\mu,T)=-\frac{\mu_0 e^2}{12\hbar^2}\int_\mathrm{BZ}\frac{\ud^2k}{4\pi^2}\frac{\Im m}{\pi}\int_{-\infty}^{+\infty} n_\mathrm{F}(E)(U-4V)\ud E\ \\
&=\frac{\mu_0 e^2}{12\hbar^2}\sum_{s=\pm} \int_\mathrm{BZ}\left[(U_1-V_1-4V_2)\left(n_F' -s \frac{n_F}{\varepsilon_{\bm k}}\right)+U_2 s\frac{n_F}{\varepsilon_{\bm k}}-V_1\varepsilon_{\bm k} n_F''\right]\frac{\ud^2k}{4\pi^2}
\end{align}
with $n_F$ a shorthand notation for $n_F(\varepsilon_{s \bm k})$ and $n_F'$, $n_F''$ are first and second derivatives of $n_F$.

\section{Low-energy approach}
\label{appendix:linearized_graphene}
The idea behind the \emph{low-energy approach} is to compute the susceptibility using formula~(\ref{eq:chi}), but with a simplified (linear or quadratic) Hamiltonian obtained in the vicinity of an energy of interest (for example a band touching point). The graphene Hamiltonian is approximated by the linearised massless Dirac Hamiltonian (for a single valley). This approximation is expected to be true for vanishing chemical potential only. Thus, $h\approx\tilde{\bm f}_\mathrm{gr}\cdot\bm\sigma$ with $\tilde{\bm f}_\mathrm{gr}$ the linear approximation of $\bm f_\mathrm{gr}$ near a Dirac point $K(0,4\pi/(3\sqrt3))$. Writing $\bm k=\bm k_\mathrm{D}+\bm q$, the approximate Hamiltonian is $h\approx q_x\sigma_x+k_y\sigma_y$.

The second term of Eq.~(\ref{eq:chi}) vanishes because of the linearity of the spectrum. The approximate Green's function yields
\begin{equation}
g=\frac{1}{E-h}=\frac{E+\bm f_\mathrm{gr}\cdot\bm\sigma}{E^2-\varepsilon^2}\approx\frac{E+q_x\sigma_x+q_y\sigma_y}{E^2-q^2}
\end{equation}
such that
\begin{equation}
\Tr\{h^xgh^ygh^xgh^yg\}=\frac{16q_x^2q_y^2}{(E-q^2)^4}-\frac{2}{(E^2-q^2)^2}
\end{equation}

To compute the double integral on $q_x$ and $q_y$, polar coordinates $q$ and $\theta$ ($\bm q=qe^{i\theta}$) are more adequate:
\begin{equation}
\int \Tr\{h^xgh^ygh^xgh^yg\}\frac{\ud^2q}{4\pi^2}=\int_0^\infty q\ud q\left[\frac{16q^2q^2}{(E-q^2)^4}\int_0^{2\pi}\cos^2\theta\sin^2\theta\ud\theta-\frac{2}{(E^2-q^2)^2}2\pi\right].
\end{equation}

With a partial fraction decomposition of $1/(E^2-q^2)^n$ with $n=2,4$, the integral over $q$ can be performed:
\begin{equation}
\int \Tr\{h^xgh^ygh^xgh^yg\}\frac{\ud^2q}{4\pi^2}=\frac{1}{3\pi E^2}
\end{equation}

Finally, the integral over the energy $E$ is computed using Eq.~(\ref{eq:integral_E}):
\begin{equation}
\chi_\mathrm{orb}(\mu,T)=\frac{3\chi_0}{4\pi}n_\mathrm{F}'(0)\stackrel{T\to0}{\longrightarrow}-\frac{3\chi_0}{4\pi}\delta(\mu)
\end{equation}
taking into account the two valleys. This is the result of Eq.~(\ref{eq:chi_McClure}). Note, however, that this approach does not capture the paramagnetic plateau $\chi_\textrm{pl}\approx 0.089 \chi_0$. Indeed, the correct result in the vicinity of $\mu=0$ is $\chi_\textrm{pl}+\frac{3\chi_0}{4\pi}n_\mathrm{F}'(0)$. This shows that a low-energy approach, that includes band coupling, \emph{i.e.} the spinor structure of the massless Dirac wavefunction, in the vicinity of $\mu=0$ does not fully recover the correct orbital susceptibility. This proves that the latter is not a Fermi surface property only but also depends on all the filled bands.

\end{widetext}

\end{document}